\documentclass[preprint]{aastex}
\usepackage{emulateapj5}

\newcommand{\uv}{\mbox{$u$-$v$}}
\newcommand{\ex}[1]{\mbox{$\times 10^{#1}$}}

\newcommand{\Msol}{\mbox{$M_\sun$}}

\newcommand{\kms}{\mbox{km s$^{-1}$}}
\newcommand{\muas}{\mbox{$\mu$as}}

\newcommand{\Jpb}{\mbox Jy~beam$^{-1}$}
\newcommand{\Ra}[4]{\mbox{${#1}^{\rm h} \; {#2}^{\rm m} \; {#3}\fs{#4} $}}
\newcommand{\dec}[4]{\mbox{${#1}\arcdeg \; {#2}\arcmin \; {#3}\farcs{#4} $}}

\newcommand{\rhoCSM}{\mbox{$\rho_{\rm CSM}$}}

\newcommand{\thout}{\mbox{$\theta_{\rm o}$}}
\newcommand{\thin}{\mbox{$\theta_{\rm i}$}}
\newcommand{\rfl}{\mbox{$r_{\rm90\%\;flux}$}}
\newcommand{\thfl}{\mbox{$\theta_{\rm90\%\;flux}$}}
\newcommand{\thpk}{\mbox{$\theta_{\rm10\%\;peak}$}}

\shortauthors{Bietenholz, Bartel \& Rupen}
\shorttitle{SN~1986J VLBI}

\begin{document}

\title{SN 1986J VLBI. The Evolution and Deceleration of the Complex
Source and a Search for a Pulsar Nebula}

\author{M. F. Bietenholz and  N. Bartel \\
Department of Physics and Astronomy, York University, Toronto, 
M3J~1P3, Ontario, Canada \\
and \\
M. P. Rupen \\
National Radio Astronomy Observatory, Socorro, New Mexico 
87801, USA}
  
\begin{abstract}

We report on VLBI observations of supernova 1986J in the spiral galaxy
NGC~891 at two new epochs, 1990 July and 1999 February, $t=7.4$ and
15.9 yr after the explosion, and on a comprehensive analysis of these
and earlier observations from $t\sim 4$ yr after the explosion date,
which we estimate to be $1983.2\pm 1.1$. The source is a shell or
composite, and continues to show a complex morphology with large
brightness modulations along the ridge and with protrusions.  The
supernova is moderately to strongly decelerated.  The average outer
radius expands as $t\,^{0.71 \pm 0.11}$, and the expansion velocity
has slowed to 6000~\kms\ at $t=15.9$~yr from an extrapolated
20,000~\kms\ at $t=0.25$~yr.  The structure changes significantly with
time, showing that the evolution is not self-similar.  The shell
structure is best visible at the latest epoch, when the protrusions
have diminished somewhat in prominence and a new, compact component
has appeared.
The radio spectrum shows a clear inversion above 10~GHz.  This might be
related to a pulsar nebula becoming visible through the debris of the
explosion.  The radio flux density between 1.5 and 23~GHz decreases
strongly with time, with the flux density $\propto t^{-2.94\pm0.24}$
between $t \sim 15$ to 19~yr.  This decrease is much more rapid than
that found in earlier measurements up to $t \sim 6$~yr.

\end{abstract}


\section{Introduction}

SN~1986J was discovered at a radio frequency of 1.4 GHz in the edge-on
spiral galaxy NGC~891 south south-west of the galaxy's center by van
Gorkom et~al.\ (1986; see also Rupen et~al.\ 1987) on 1986 August
21/22.  The distance to NGC~891 was determined to be $\sim 10$~Mpc
\citep[see e.g.][]{Tonry+2001, Ferrarese+2000, Tully1988,
KraanKorteweg1986, Aaronson1982}, and we will adopt a round value of
10~Mpc throughout this paper.  With a peak flux density at 5~GHz of
128 mJy \citep[][WPS90 hereafter]{WeilerPS1990}, SN~1986J is
one of the radio-brightest supernovae ever
detected. Very-long-baseline interferometry (VLBI) observations were
made soon after the discovery and revealed an elongated brightness
distribution \citep{BartelSR1989}. Shortly after, in 1988 September 29
(1988.7), further observations using a more sensitive VLBI array
allowed an image to be made, the first one of any optical supernova
\citep[ B91 hereafter]{RupenBF1991, Bartel+1991}. The image showed a
complex source, perhaps with a composite structure, and a marginal
indication of a shell with a highly modulated brightness distribution
along the ridge and with at least one protrusion directed to the
south-east and another one directed to the north-west.

Optical observations, partly made even before the radio discovery,
showed a faint object of $\sim 18$ mag that decayed unusually slowly
\citep{Rupen+1987}. Prominent H$\alpha$ lines in the spectrum led to
the classification as a type~II supernova. The spectral lines, however,
were surprisingly narrow, with a full-width at half-maximum (FWHM) of
$\lesssim 1000$~\kms\ \citep{Rupen+1987, Leibundgut+1991}, at
least an order of magnitude smaller than the expansion velocity
expected for the shock front.  Interestingly, similarly narrow lines
were found for SN~1988Z \citep{StathakisS1991}, but in addition dim
and much broader lines were also present at early times for that
supernova.  Perhaps such broad lines also existed for SN~1986J, but
had already become too dim at the time of discovery. A large Balmer
decrement and a small extinction together with ``metallic'' lines were
also found for SN~1986J, which led \citet{Rupen+1987} to propose a
very high electron density of $n_e > 10^9$~cm$^{-3}$ in the emitting
region, along with regions of much lower
density. \citet{Chevalier1987} interpreted the slow decay of the
optical emission as being due to energy input from a central pulsar
and the narrow H$\alpha$ emission-lines as originating in the central
region of the supernova.  A different interpretation of the narrow
H$\alpha$ emission-lines is offered by \citet{Chugai1993}, who does
not relate them to an inner emission zone, but rather to shock-excited
dense clouds of gas in the circumstellar material, which move much
more slowly than the shock front of the supernova.

The explosion date of SN~1986J is not well known.  The earliest
pre-discovery radio detection was on 1984 May 1
\citep{vdHulstdBA1986}.  Considering this detection along with
subsequent measurements, \citet{Rupen+1987} concluded that SN~1986J
probably exploded around 1982 to 1983.  \citet{Chevalier1987} used a
more extended set of flux density measurements and fit his widely-used
circumstellar interaction or mini-shell model \citep{Chevalier1982a,
Chevalier1982b} to the data and obtained an explosion date of
$1983.0\pm 0.5$.  Using a complete set of radio flux density
measurements at five frequencies up to 1988 December 28, WPS90
obtained an explosion date of $1982.7^{+0.8}_{-0.5}$.

The progenitor was believed to have been a red supergiant with a mass
of $\sim 20$ to 60~\Msol\ \citep{Rupen+1987} or $\sim 20$ to 30~\Msol\
(WPS90) and estimated to have rapidly lost mass into a clumpy
circumstellar medium (CSM) with a mass-loss to wind-velocity ratio of
${\dot M / w} \sim 2.4\times 10^{-4}$~\Msol~yr$^{-1}$ per 10~\kms\
(WPS90; note that this value is somewhat dependent on assumptions made
by those authors).

The canonical interpretation of the radio emission from supernovae is
that it is synchrotron radiation produced in the region around the
contact discontinuity where the supernova ejecta interact with the
CSM\@.  The evolution of the expanding radio shell provides an
observational window on the structure of both the CSM and the
supernova ejecta, and is therefore of considerable interest.  The
interaction region is bounded on the outside by the forward shock that
travels outward from the contact discontinuity into the CSM, and on
the inside by the reverse shock that travels, in co-moving
coordinates, back into the ejecta. In his mini-shell model of an
expanding supernova, Chevalier approximated the mass density profiles
of the CSM and of the ejecta by power laws in radius, $r$, with the
density of the CSM being $\propto r^{-s}$ and that of the ejecta being
$\propto r^{-n}$.  In this case, self-similar solutions can be found,
and the supernova expands with $r \propto t^m$, with the deceleration
parameter $m$ given by $m= (n-3) / (n-s)$.  The ratio between the
radius of the forward shock, or the outer radius of the radio shell, and
the radius of the reverse shock, or the inner radius of the radio shell,
defines the shell thickness.  In the case of self-similar expansion as
given by the Chevalier model, this ratio remains constant, and is
thought be 1.21 to 1.29 for $20 > n > 7$.

At early times after shock breakout, most of the radio radiation from
the interaction region is free-free absorbed by the photo-ionized CSM,
with the optical depth, $\tau$, being a function of $\dot M/w$ of the
progenitor.  As the shock front expands, it sweeps away the obscuring
layers of the CSM, decreasing the external absorption and causing the
flux density to rise quickly.  After the absorption has become small
and the radio shell optically thin, 
the flux density, $S_\nu$, at frequency $\nu$, decays as $\propto
t^\beta$, where $\beta$ is related to the radio spectral index,
$\alpha$ ($S_{\nu}\propto \nu^{\alpha}$), and to $m$ by $\beta =
\alpha-3+3m$ \citep{Chevalier1982b}.

In the case of SN~1986J, the flux densities rose more slowly than
predicted by this model.  This was accounted for by assuming
additional absorption within the radio shell itself.  With this
extension to the model, the rise of the radio light curves could be
well fit (Chevalier 1987; WPS90).  An unstable shock front or
filamentation in the ejecta (WPS90) might well produce such mixed
absorption.

This model and its extensions have generally been successful in
describing the early evolution of radio supernovae.  However, caution
is probably warranted in applying it to SN~1986J, which is a complex
source (B91).  Furthermore, deviations from a self-similar evolution
have been found for SN~1993J \citep{Bartel+2002, Bartel+2000}, and
could also be expected for SN~1986J\@. Thus a hydrodynamic model
\citep[e.g.,][]{MioduszewskiDB2001} for the evolution of a supernova
shell may be needed. High-resolution measurements of the evolution of
the radio supernova morphology and the deceleration of the expansion
are therefore of particular interest to study the dynamics of the
interaction between the ejecta and the CSM, and to guide the
development of such more elaborate models.  Furthermore, and perhaps
most importantly, such measurements may reveal a young pulsar nebula
surrounding the compact remnant of the explosion.

Here we report on new VLBI observations of SN~1986J at two further
epochs, 1990 July 21 (1990.6) and 1999 February 22 (1999.1), and on a
comprehensive analysis of these and earlier observations from 1987
February 23 onward.
In \S\ref{obss} we describe our VLBI and VLA observations and data
reduction.  We then give our VLBI results. In \S\ref{imagess} we
present a high-resolution image from our latest epoch of observations,
and give a comprehensive analysis of the images from three epochs from
1988.7 onward.
In \S\ref{lightcurves} we give flux density measurements from 1998 to
2002, and compare them with the radio light curve deduced from earlier
measurements.  We also derive the radio spectrum of SN~1986J, which
exhibits a high-frequency inversion. In \S\ref{astroms} we give our
astrometric results, which will likely be of importance for future
detailed studies of the dynamical evolution of SN~1986J\@. In
\S\ref{sizes} we determine the size of SN~1986J at different epochs, and
in \S\ref{decels} the deceleration of the expansion. In
\S\ref{bfields} we infer the evolution of the magnetic field, and
finally, in \S\ref{discuss} we discuss our results and in
\S\ref{concs} give our conclusions.

\section{Observations and Data Reduction\label{obss}}

\subsection{VLBI Measurements \label{vlbiobss}}

The VLBI observations were made with global VLBI arrays of 8 and 14
antennas with a duration of 12 to 22 hours for each run. The
declination of +42\arcdeg\ of SN~1986J enabled us to obtain dense and
only moderately elliptical \uv~coverage.  As usual, a hydrogen maser
was used as a time and frequency standard at each telescope. The data
were recorded with the MKIII and either the VLBA or the MKIV VLBI
systems with sampling rates of 112 or 256~Mbits per second.  We
observed at frequencies of 8.4~GHz on 1988 September 29 and 1990 July
21, and at 5.0~GHz on 1999 February 22.  The characteristics of the
observations are given in Table~\ref{obst}.

The data from the first two sessions were correlated with the MKIII
VLBI processor at Haystack observatory, and the data from the last
session with the NRAO VLBA processor at Socorro.  The analysis was
carried out with NRAO's Astronomical Image Processing System (AIPS).
The initial flux density calibration was done through measurements of
the system temperature at each telescope.

In all three sessions we observed, besides SN~1986J, the nearby compact
source, 3C~66A, which is $\sim 40\arcmin$ away from SN~1986J,
and used it to calibrate the variation of the amplitude gains with
time.  For the first two observations the brightness of the supernova
was still sufficient for obtaining detections for each of the
two-element interferometers of the array during a scan time of a few
minutes. At the time of our last VLBI observations, SN~1986J was too
faint for obtaining such detections, and accordingly we observed by
phase-referencing to 3C~66A, using a cycle time of $\sim 3.5$~min with
2~min spent on SN~1986J and $\sim 1$~min on 3C~66A.
 
Typically in VLBI the antenna phases are essentially unknown, and are
determined by self-calibration, starting with a point-source model
\citep{Walker1999}.  This procedure usually converges well, but it can
introduce symmetrizing and other artifacts into the images
\citep{MassiA1999, Linfield1986}.  Phase-referencing to a source of
known structure allows the antenna phases to be determined from the
calibrator source, and hence will allow the most un-biased images of
the target source.  In addition, phase-referencing also allows the
determination of an accurate position for the target source relative
to that of the reference source.

For each observation we first fringe-fit 3C~66A, and then in an
iterative procedure self-calibrated and imaged it to determine the
complex antenna gains as a function of time.  These antenna gains,
which are corrected for the source structure of 3C~66A as a result of
using images of it in the self-calibration, were then interpolated and
applied to SN~1986J\@.  For the first two observations, we used only the
amplitude portion of the gains derived from 3C~66A, and SN~1986J was
fringe-fit and then self-calibrated, starting with point-source
models and progressing to CLEAN component models.  For the 1999.1
epoch, both the phases and amplitudes were interpolated and applied to
SN~1986J and a phase-referenced image was made.  In addition to
imaging, as one means of consistently estimating the size of
SN~1986J at our three different epochs, we also fit a geometrical
model directly to the calibrated visibility data by weighted least
squares.

\subsection{VLA Measurements \label{vlas}}

In addition to using the phased VLA as an element in our VLBI array,
we obtained interferometric data with the VLA on 1998 June 5 (1998.4),
concurrently with the VLBI run on 1999 February 22 (1999.1), and in
2002 May 25 (2002.4).  In each of the VLA 
sessions we observed in both senses of circular polarization, and used
a bandwidth per IF of 50~MHz. 
In 1998.4, the VLA was in the AnB configuration, and we observed at
1.4, 5.0, and 8.4~GHz.  No secondary phase calibrator was used for
this session, and the amplitudes were calibrated\footnote{This session
was scheduled {\em ad hoc}\/, and we used at each frequency only that
selection of antennas for which 3C~48 could reasonably be approximated
as a point source.  At 8.4~GHz, this amounted to using only half of
the antennas.  The flux density, however, is still well determined,
since even with this limitation, the statistical uncertainties remain
small compared to the systematic ones, which are not directly
dependent on the number of antennas.}
w.r.t.\ 3C~48.  In 1999.1, the VLA was in the CnD configuration, and
we observed at 1.4, 5.0, 8.4, 14.5, and 22.5~GHz.  We used the
amplitude calibrator, 3C~286, and the secondary phase calibrator
0230+405.  In 2002.4 the VLA was in the AnB configuration and we
observed at 1.4, 5.0, 8.4, 14.5, 22.5, and 43.3~GHz.  We used the
amplitude calibrator 3C~48, except at 43~GHz which case is discussed
separately below, and the secondary phase calibrator 0303+472.

The data at frequencies below 43~GHz were reduced following standard
procedures, with the amplitudes calibrated by using observations of
the standard flux density calibrators (3C~286 and 3C~48) on the scale
of \citet{Baars+1977}.  The changes in gain between the program
sources and the flux calibrators were $< 2$\%. A short timescale of
20~s was used for the phase calibration of the calibrator sources to
avoid any possible decorrelation caused by rapid phase
fluctuations.  The amplitude calibrators were at elevations comparable
to our program sources. At frequencies above 10~GHz, we applied an
atmospheric absorption correction based on an average zenith
absorption, in Nepers, of 0.02 at 14.5~GHz, and 0.05 at 23.4~GHz.
The atmospheric absorption is dependent on the weather at the time of
observations.  However, the average zenith opacity for our observing
frequencies is not large.  Since the zenith distances for our
observations were also not large, the corrections, and any errors in
them, were small. In particular, the change in gain between the flux
calibrator and program source due to absorption was $< 1$\%, so
typical deviations of the zenith absorption from our average values
would not significantly affect our results.

At 43~GHz, the recommended flux calibrator is 0713+438.  Due to
scheduling constraints, we could observe this source only at a low
elevation.  We therefore also used another source, 3C~84, as an
amplitude calibrator.  The 43-GHz flux density of 3C~84 was $6.7 \pm
0.7$~Jy in 2002 June (C. Chandler; private communication), and it was
observed at an elevation comparable to that of SN~1986J\@.  We applied
an atmospheric opacity correction based on an average zenith opacity
of 0.07 Nepers.  In addition, at this frequency, we corrected for the
individual gain curves of the VLA antennas as measured in 2001.

\begin{figure}
\epsscale{0.75}
\plotone{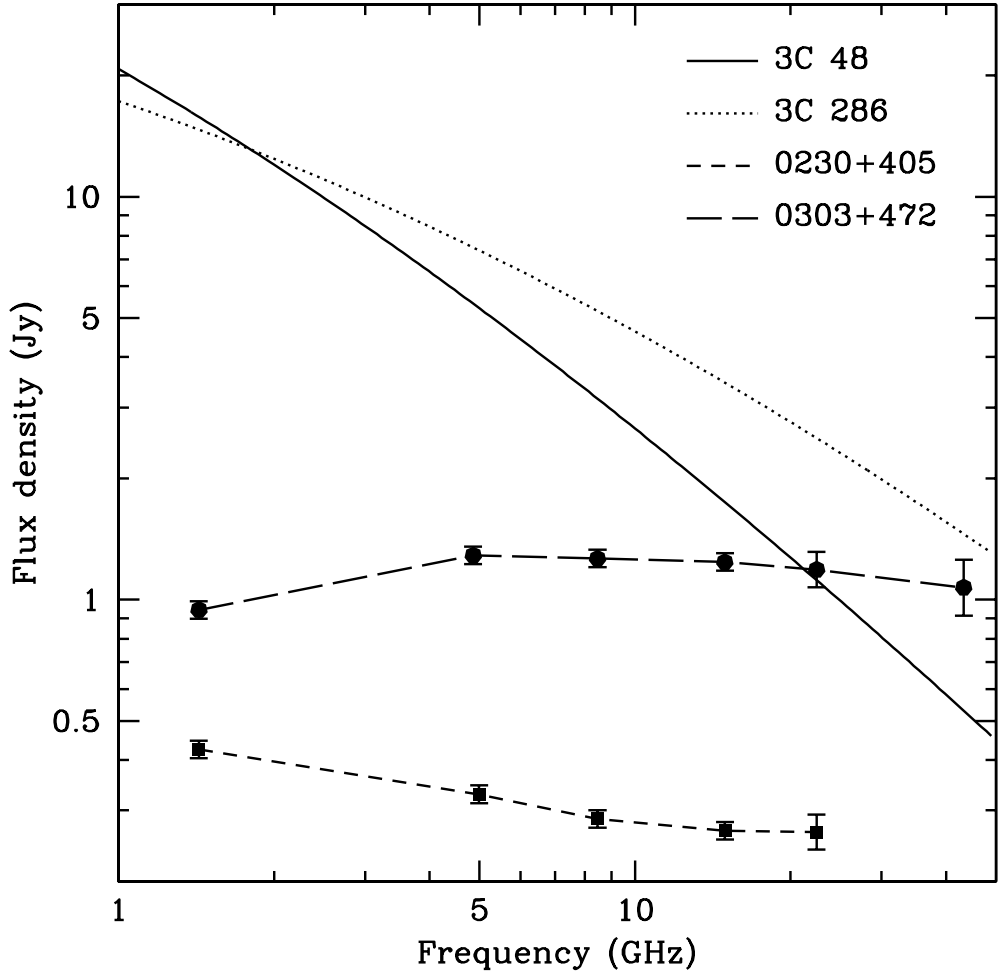}
\figcaption{The radio spectra of the calibrator sources used for the
VLA flux density observations.  On 1999 February 22, the flux
calibrator was 3C~286 and the secondary calibrator was 0230+405.  On
2002 May 24 The flux calibrator was 3C~48 (except at 43~GHz, see text)
and the secondary calibrator was 0303+472.  We plot the adopted
polynomial spectra for 3C~48 and 3C~286.  The coefficients, as implemented
in modern versions of AIPS, are (Perley, private communication) as follows,
where $S_\nu$ is in Jy and $x$ is log($\nu$ in GHz). For
3C~286, $S_\nu = 1.23734 - 0.43276x - 0.14223x^2 + 0.00345x^3$, and
for 3C~48, $S_\nu = 1.31752 - 0.74090x - 0.16708x^2 + 0.01525x^3$.
\label{vlacalf}}
\end{figure}

We plot the polynomial spectra adopted for 3C~48 and 3C~286
in Figure~\ref{vlacalf}.  Also in this figure, we plot the derived
spectra of the two phase calibrator sources, 0230+405 and 0303+472,
along with their uncertainties, which are dominated by the systematic
uncertainty in the VLA amplitude calibration.

\begin{figure}
\epsscale{0.8}
\plotone{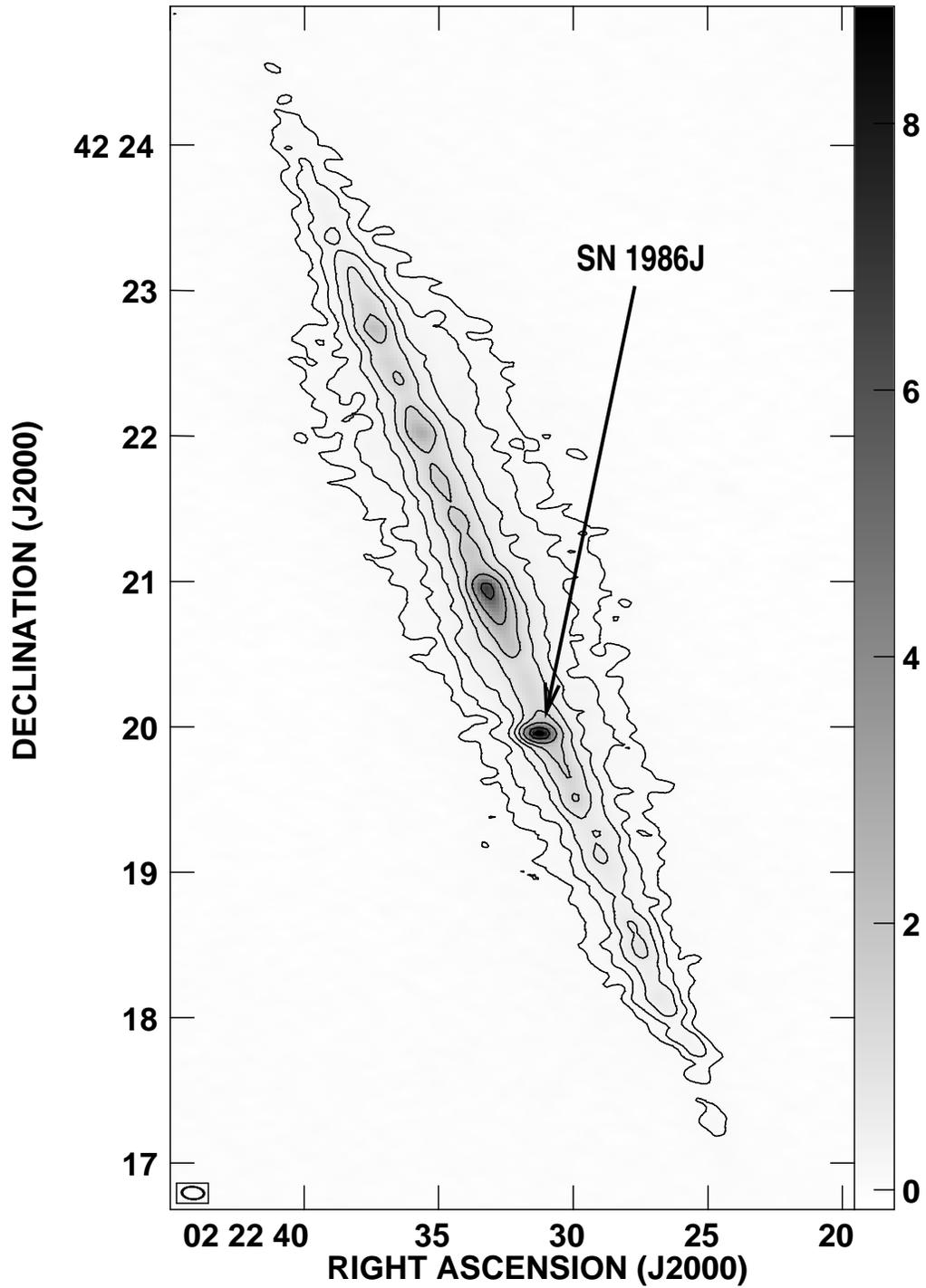}
\figcaption{A VLA image of NCC~891 at 5.0~GHz on 1999 Feb.\ 22.  The
contours are drawn at 1, 2, 4, \dots, 64, and 90\% of the peak
brightness of 8.9~m\Jpb.  The grayscale is labeled in m\Jpb. The rms
of the background brightness was 33~$\mu$\Jpb. The CLEAN beam is
plotted at lower left, and its FWHM was $9.1\arcsec \times 5.7\arcsec$
at p.a.\ 86\arcdeg.  SN~1986J is indicated.
\label{vlaimagef}}
\end{figure}

We imaged the radio emission of the galaxy NGC~891 at 5.0~GHz in
1999.1, and show the result in Figure~\ref{vlaimagef}.  SN~1986J is
located south-west of the galactic nucleus.  We also obtained total
flux densities of SN~1986J at several frequencies at each epoch.  With
the exception of the one at 43~GHz, these flux densities were derived
from data which were self-calibrated in phase but not in amplitude.

For the standard error of the total flux density measurements, we take
the sum in quadrature of the statistical standard error, an
uncertainty in the VLA flux density calibration, and the additional
systematic uncertainties detailed below.  The uncertainty in the VLA
flux density calibration was taken to be 5\% for frequencies up to
14.5~GHz, 10\% at 23.5~GHz, and 14\% at 43~GHz (we use an uncertainty
somewhat higher than usual at 43~GHz because of the aforementioned
difficulty in amplitude calibration for that frequency).  An
additional systematic uncertainty arises because confusion needs to be
taken into account for frequencies below 5~GHz at epoch 1999.1, for
which the VLA was in the CnD configuration and the resolution was
moderate.  In particular, for these frequencies, we derive our flux
densities from a combination of model-fits in the image plane and
\uv~plane, with the contribution from the galactic core and disk of
NGC~891 accounted for in the model-fits.  In 1999.1 the brightness of
NGC~891 was $\sim 2$ and $\sim 0.15$ times the peak brightness of
SN~1986J at 1.5 and 5~GHz, respectively.  The additional systematic
uncertainty for these measurements is computed from the ambiguity of
these model-fits.  At 43~GHz, we take an additional systematic upward
uncertainty of 50\% because we could not self-calibrate in phase, and
so may have underestimated the true flux density.

\section{VLBI Images of SN~1986J\label{imagess}}
\subsection{A High-Resolution Image at the Latest Epoch}

\begin{figure}
\epsscale{0.80}
\plotone{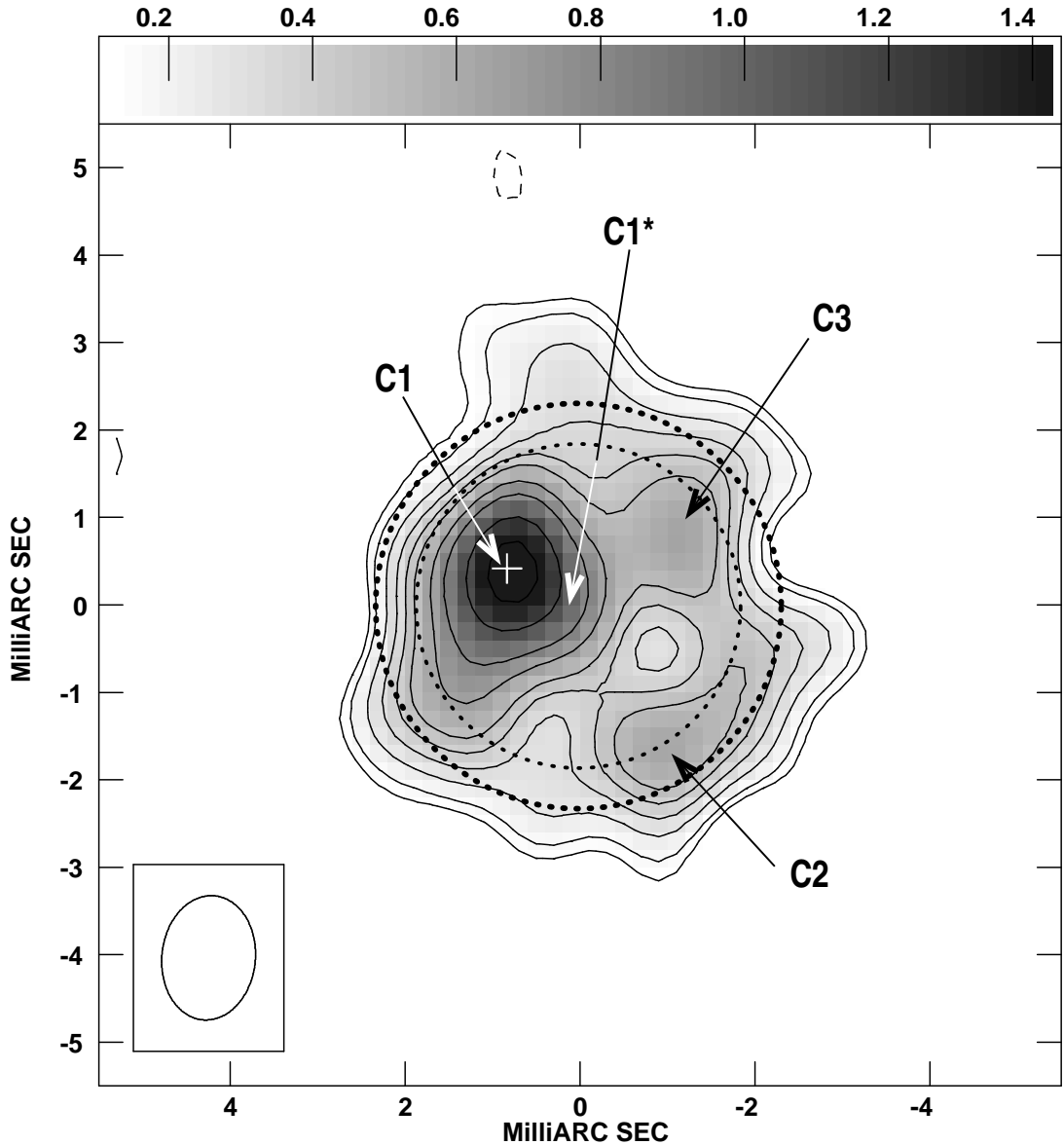}
\figcaption{A high-resolution image of SN~1986J on 1999 February 22 at
5~GHz.
The CLEAN beam was 1.42~mas $\times$ 1.07~mas FWHM at p.a.\ $=
-5$\arcdeg\ and is plotted at lower left.  We chose the CLEAN beam to
have the same area, but be somewhat more circular than an elliptical
Gaussian fit to the inner portion of the dirty beam.  Contours are
drawn at $-7.4$, 7.4, 10, 15, 20, 25, 30, 40, 50, 70, and 90\% of the
peak brightness of 1.64~m\Jpb. The rms of the background brightness
was 40~$\mu$\Jpb.  The dotted circles indicate the inner and outer
radii of a spherical shell fit to the \uv~data (see text,
\S\ref{uvfits}).  Note that the brightest portion of a projected shell
of uniform volume emissivity is at the inner radius. The white cross
indicates the position of the point source also fit to the data.
North is up, and east is to the left.
\label{hiresf}}
\end{figure}

In Figure~\ref{hiresf}, we show a 5-GHz image of SN~1986J at epoch
1999.1 made from our most sensitive observations using
phase-referenced imaging.  The image shows a rather complex extended
radio source.  The outer contours are very roughly circular, but the
brightest point lies to the north-east of its apparent center.  There
is a minimum in the interior, which lies somewhat to the south-west of
the apparent center.  This structure could be interpreted as a shell,
albeit a quite heavily modulated one.  An optically thin, spherical
shell of uniform volume emissivity would have a brightness
distribution with a minimum in the center and a circular ridge-line
which is close to its projected inner radius.  The most prominent
deviation of our image of SN~1986J from such a geometry is the
prominent brightness peak to the north-east, which we label C1.  The
peak flux density per beam at the location of C1 is $\sim 1.6$~mJy.

As mentioned before, we also fit a geometrical model directly to the
\uv~data.  Our choice of model was motivated by the morphology of the
above image, and our model consists of a spherical shell of uniform
volume emissivity and a point source to represent C1.  We defer the
detailed discussion of these model fits to \S\ref{uvfits} below.
However, we indicate in Figure~\ref{hiresf} the inner and outer radii
of the fit shell, and we take the center of the fit shell as origin of
the coordinate system.  The fit point source is also indicated on the
figure. It has a flux density of $\sim 1$~mJy, which indicates that C1
is approximately twice as bright as the shell emission at that
location.  We can then estimate C1 to have $\sim 16$\% of the total
flux density in the image, and to have a brightness temperature
$\gtrsim 3 \times 10^7$~K.

It is clear that the brightness distribution of SN~1986J does not
resemble that of the described model in detail. Even allowing for the
presence of C1, the shell brightness is heavily modulated.  Two other
components are visible, a south-western component C2 and a
north-north-western component C3, each having a brightness a third 
that of C1. They can be interpreted as being part of the ridge of the
shell.  In addition to the shell, and to C1, C2 and C3, there are
protrusions which add to the complexity of the image and cause
significant deviations from circularity of the outer contours.

C1 itself, while relatively compact, may well not be entirely
pointlike.  Comparing its contours to those of the CLEAN beam shows an
extension both toward the geometric center, labeled C1*, and toward
the inner radius of the fit shell to the south. While part of the
emission of C1, in particular the extension to the south, can probably
also be interpreted as being part of the ridge, the largest part of C1
is located rather further toward the geometric center than one might
expect from a shell component, unless it is an unusually strong
component located on the front of the shell.
 
Qualitatively, the structure of the supernova is similar to that found
in 1988.7 by B91. The shell structure, only marginally discernible in
the 1988.7 image, has now become somewhat clearer.  A north-eastern
component dominated the image at the early epoch and such a component
still dominates the image in 1999.1.  Protrusions distorted the image
in 1988.7 and still characterize the appearance of the supernova in
1999.1. Apart from these general similarities however, clear
differences in the structure of the supernova have developed over the
years.

\subsection{A Comparative Study of the Sequence of Images at the Three Epochs}

To investigate the structural evolution of the supernova in more
detail, we produced a sequence of images of SN~1986J at three
consecutive epochs from 1988.7 to 1999.1, approximately $t= 6$, 7, and
16~yr after shock breakout.  To better search for possible changes in
the images over the years we adopt a more conservative image
reconstruction technique than that used for making the image in
Figure~\ref{hiresf}.  In particular, we adopt a weighting scheme for
the correlated flux densities which increases the dynamic range of the
images.  Since residual calibration errors may be present in the data
and need not scale with the thermal noise, we compressed the weights
somewhat by weighting by the inverse of the thermal noise rms, rather
than the usual inverse of the thermal noise variance.  Then, we choose
a beam size based on the 1999.1 image made using Briggs' robust
weighting scheme \citep{BriggsSS1999, Briggs1995}, with the robustness
parameter in AIPS set to zero to give the best trade-off between
signal-to-noise ratio and minimized sidelobes.  A Gaussian fit to the
inner portion of the dirty beam produced with this weighting gives a
CLEAN beam with a full-width at half-maximum (FWHM) of 2.06~mas
$\times$ 1.04~mas at a p.a.\ of $-7$\arcdeg.

The angular resolution of our observations were different for each
epoch.  Therefore, in order to minimize any bias in the comparison of
the images that could be introduced by the convolution with an
inconsistently changing CLEAN beam, we choose to convolve our three
images with a CLEAN beam whose size evolves similarly to that of the
supernova. In particular we choose a CLEAN beam whose size evolves
as $t^{0.75}$, in anticipation of our results on the deceleration
of the supernova expansion described in \S\ref{decels} below. In order
to accommodate the somewhat different angles of elongation of the
original beams, we increase the length of the minor axis of the beam
so that in 1999.1 it is 1.40~mas.  With such a dynamical CLEAN beam,
the appearance of the supernova would essentially not
change apart from a scaling factor if the supernova evolved
self-similarly.

We show the sequence of images produced in this way in
Figure~\ref{imagesf}. It clearly shows that the supernova is
expanding.
This sequence is only the second, after that for SN~1993J
\citep[e.g.,][]{Bietenholz+2002, Bartel+2000, Rupen+1998,
Marcaide+1997}, where the expansion of the supernova is so clearly
visible.  It also shows that the structure is changing with time.

We now follow the evolution of the structure in more detail.  In the
1988.7 image, a strong component is located approximately in the
center of the radio source.  Comparing this component to the 50\%
contour of the CLEAN beam, it is clear that this component is
significantly extended. Judging from the early high-resolution image
of B91, it is a combination of the strong northern component and the
shell emission, since a spherical shell would be barely resolved at
the relatively coarse angular resolution we now use.  Further,
protrusions to the south-east and to the north-west are clearly
visible. B91 note parenthetically that in their high-resolution image
a faint, additional protrusion to the south-west might perhaps be
present, but we cannot confirm it in our image.  To determine the
extent to which the morphology is dependent upon the self-calibration,
we ``self''-calibrated the image from 1988 using a scaled version of the
1999 image, and found no change in the gross morphology described
above.

In 1990.6, the bright component still dominates the image. It is again
clearly extended, and presumably again a combination of the strong
northern component mentioned above and a contribution from the shell.
In addition a second component to the south-east has developed. We
think it could be component C2 and part of the ridge.  The protrusion
to the north-west is still visible. A protrusion to the south-east is
also visible, but appears to have an orientation
30\arcdeg\ further to the south than the protrusion seen in 1988.7.
It is striking that a new protrusion seems to have appeared to the
north-east.

By 1999.1, the bright component, now labeled C1, still dominates the
image. This time however, the component is compact.  We fit a model
including a point source to represent this component directly to the
\uv~data (\S\ref{uvfits} below).  If we subtract the fit point source
at the location of C1 from the image, we obtain the brightness
distribution shown in the fourth panel, which is indeed more similar
to the brightness distribution in the second panel.  The bright inner
part has rotated further, toward an angle of $\sim -60$\arcdeg, and
forms an arc that extends to C3 and C2 as part of the ridge, which is
also seen in the high-resolution image in Figure~\ref{hiresf}. The
north-west and south-east protrusions are still visible, but have
become less pronounced and contribute only mildly to the distortion of
the morphology.

Looking at the three images in a more comprehensive way, three main
characteristics are apparent: 1) The shell structure appears to have
become more prominent since it is not discernible in our first image
but becomes clearer in the following two images.  2) The
protrusions appear to have expanded more slowly than the rest of the
source.  In fact there is a possibility that they have a more erratic
nature evolving on a time scale shorter than a couple of years. 3)
The brightest component is clearly resolved in the first two images
and has a peak flux density per beam of only 21 and 17\% of the total
flux density of the source, respectively. In the last image, however,
it is much more compact and in addition more prominent, having a peak
flux density per beam of 29\% of the total flux density.

\subsection{A Search for Polarized Emission \label{polns}}

We also made images in Stokes parameters $Q$ and $U$ for the
observations in 1999.1.  We determined the antenna polarization
parameters from the observations of 3C~66A\@.  We did not detect any
significant linear polarization for SN~1986J\@. The $3\sigma$ upper
limit on the polarized brightness is 0.11~m\Jpb\ for a beam FWHM of
1.8~mas $\times$ 0.9~mas. This limit corresponds to a fractional
linear polarization of $< 7.5$\% for the peak of the image.

\section{The Radio Light Curves and the Spectrum\label{lightcurves}}

\begin{figure}
\epsscale{1.0}
\plotone{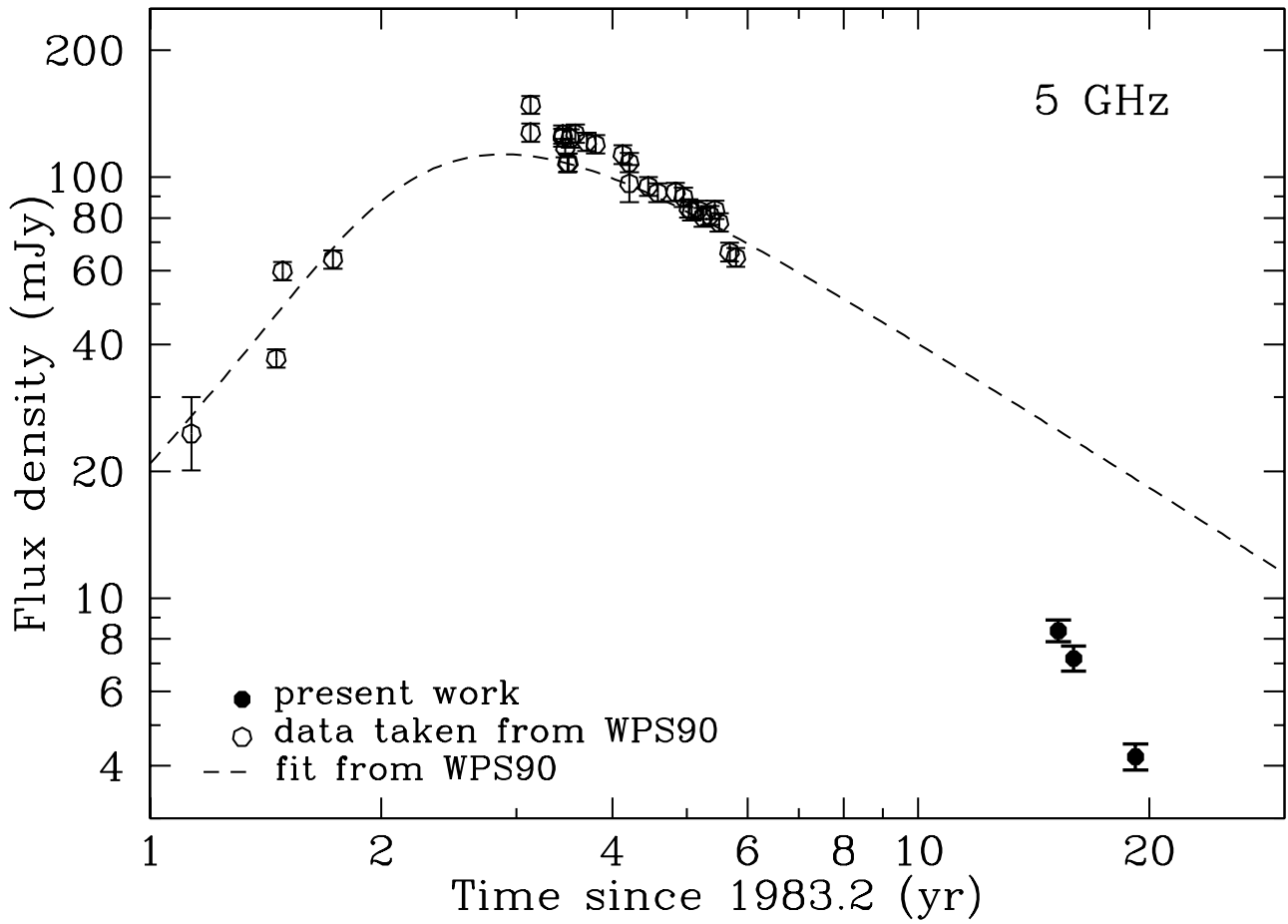}
\figcaption{The total flux density of SN~1986J as a function of time.
Our measurements are plotted as filled circles.  In addition we plot
the values taken from WPS90, along with the fit radio light curve they
derived from measurements between 0.3 and 22.5~GHz.
\label{betaf}}
\end{figure}

From interferometric VLA observations, we determined the total flux
densities at different frequencies for the epochs 1998.4, 1999.1, and
2002.4, and we list them in Table~\ref{vlafluxt}.  To illustrate the
evolution of the flux density with time, we plot our flux densities at
5~GHz, as well as those measured by WPS90, in Figure~\ref{betaf}.  The
evolution of the flux density with time, $t$, after the explosion
date, $t_0$, is usually parameterized as $S_\nu \propto (t-t_0)^\beta$
for the period after the supernova has become optically thin.  WPS90
fit a model to their flux density measurements at frequencies between
0.3 and 23~GHz, and determined that $\beta = -1.18^{+0.02}_{-0.04}$
for the period up to 1989.  We plot their fit also in
Figure~\ref{betaf}.  Our flux densities are significantly below the
extrapolation of WPS90's fit, indicating a much more rapid decay of
the flux density as a function of time.  In fact, the average value of
$\beta$ at 5~GHz between WPS90's last measurement in 1989.0 and our
first measurement in 1998.4 is $-2.22 \pm 0.08$.

\begin{figure}
\epsscale{1.0}
\plotone{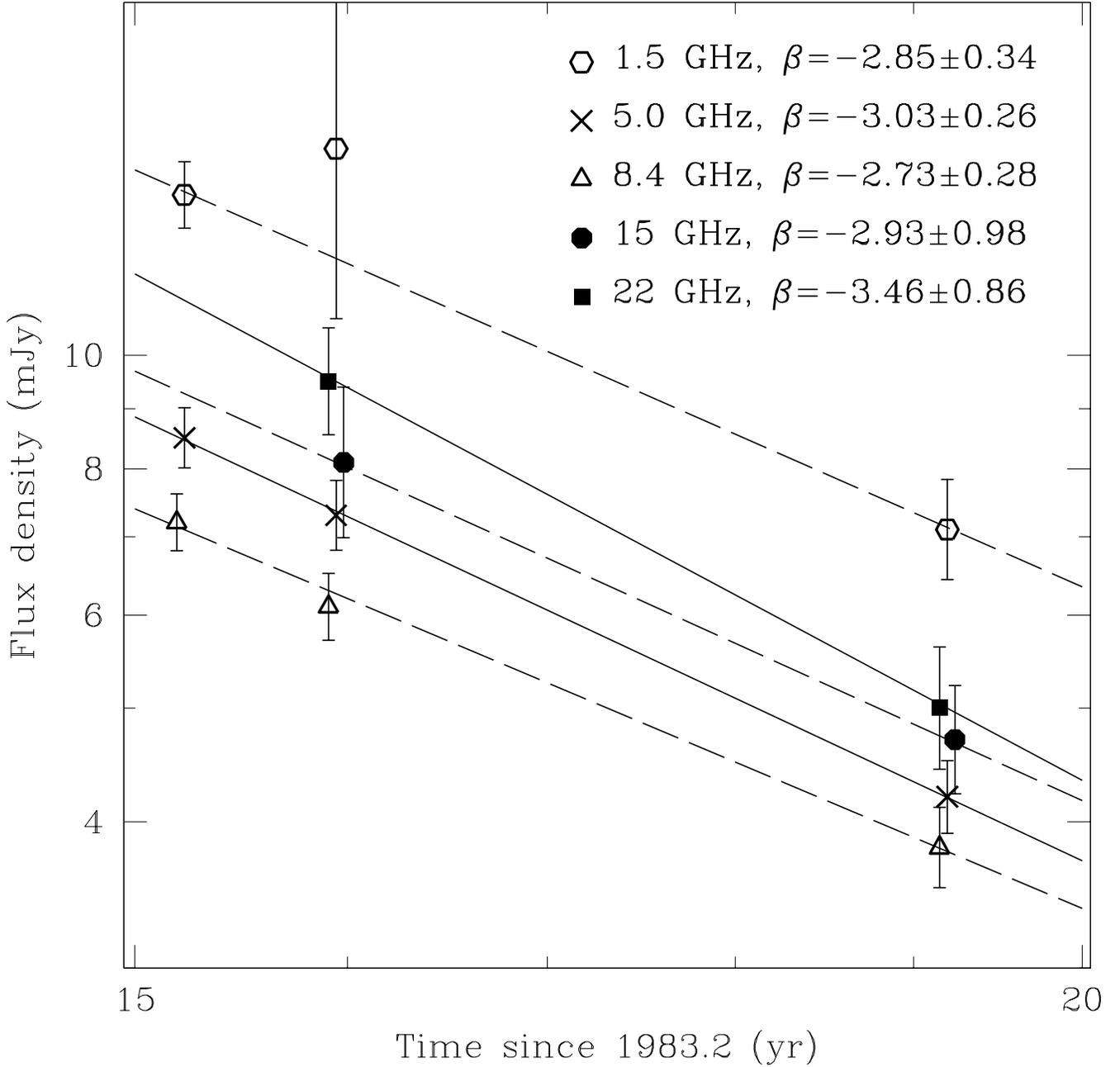}
\figcaption{The total flux density of SN~1986J as measured with the
VLA at various frequencies as a function of time.  For clarity, some
points have been shifted slightly in time.  The lines indicate
weighted fits to the flux density decay with time, with the relevant
value of $\beta$ (i.e., slope) indicated.  The standard errors include
statistical and systematic contributions.
\label{vlabetaf}}
\end{figure}

We have flux density measurements from 1998 to 2002, and therefore we
can determine the current rate of flux density decay.  We plot our
flux density measurements at 1.4 to 23~GHz as a function of time in
Figure~\ref{vlabetaf}.  We also determined $\beta_\nu$ separately at
each frequency, $\nu$, with a weighted least-squares fit, and these
fits are also indicated in the figure.  The rate of decline is even
more rapid than the average rate between 1989 and 1998, with
$\beta_\nu$ ranging from $-2.7 \pm 0.3$ at 8.4~GHz to $-3.5 \pm 0.9$
at 23~GHz.  The values of $\beta_\nu$ at our different observing
frequencies from 1.4 to 23~GHz are consistent within the
uncertainties, suggesting that between 1998 and 2002, there is no
significant change with time in the radio spectrum, and that $\beta$
is in fact independent of frequency to within our measurement errors.

The flux densities in Table~\ref{vlafluxt} show that there is a
significant inversion in the spectrum above 10~GHz.  This inversion
suggests that the spectrum can be decomposed into two components,
a ``normal'' component, with a negative spectral index, and an
``inverted'' component with a positive spectral index at least up to
23~GHz.  To consistently determine both, $\beta$, assumed now to be
independent of frequency, and the spectrum, we combined all our flux
density measurements except the one at 43~GHz.  Then we fit $\beta$
and the two power-law components of the spectrum by weighted least
squares.  We derived uncertainties on our fit parameters from a Monte
Carlo simulation, varying all the flux density measurements within
their respective uncertainties and using 4000 realizations.  We note
that this decomposition of the spectrum is probably not unique, but it
illustrates the general behavior of the spectrum and allows us to
accurately determine $\beta$.

We find that the fit value of $\beta$ is $-2.94 \pm 0.24$ between 1998
and 2002.  This value is significantly different from that measured up
to 1989, indicating that the rate of flux density decay has increased
substantially between 1989 and 1998.  An increase with time of the
rate of flux density decay has been seen for other supernovae, in
particular for 
SN~1980K \citep{Montes+1998} and for SN~1993J for which a distinct
decrease in $\beta$ from $\sim -0.7$ to $\lesssim -1.2$ was found $\sim
6.5$~yr after the explosion \citep{Bartel+2002}.

\begin{figure}
\epsscale{1.0}
\plotone{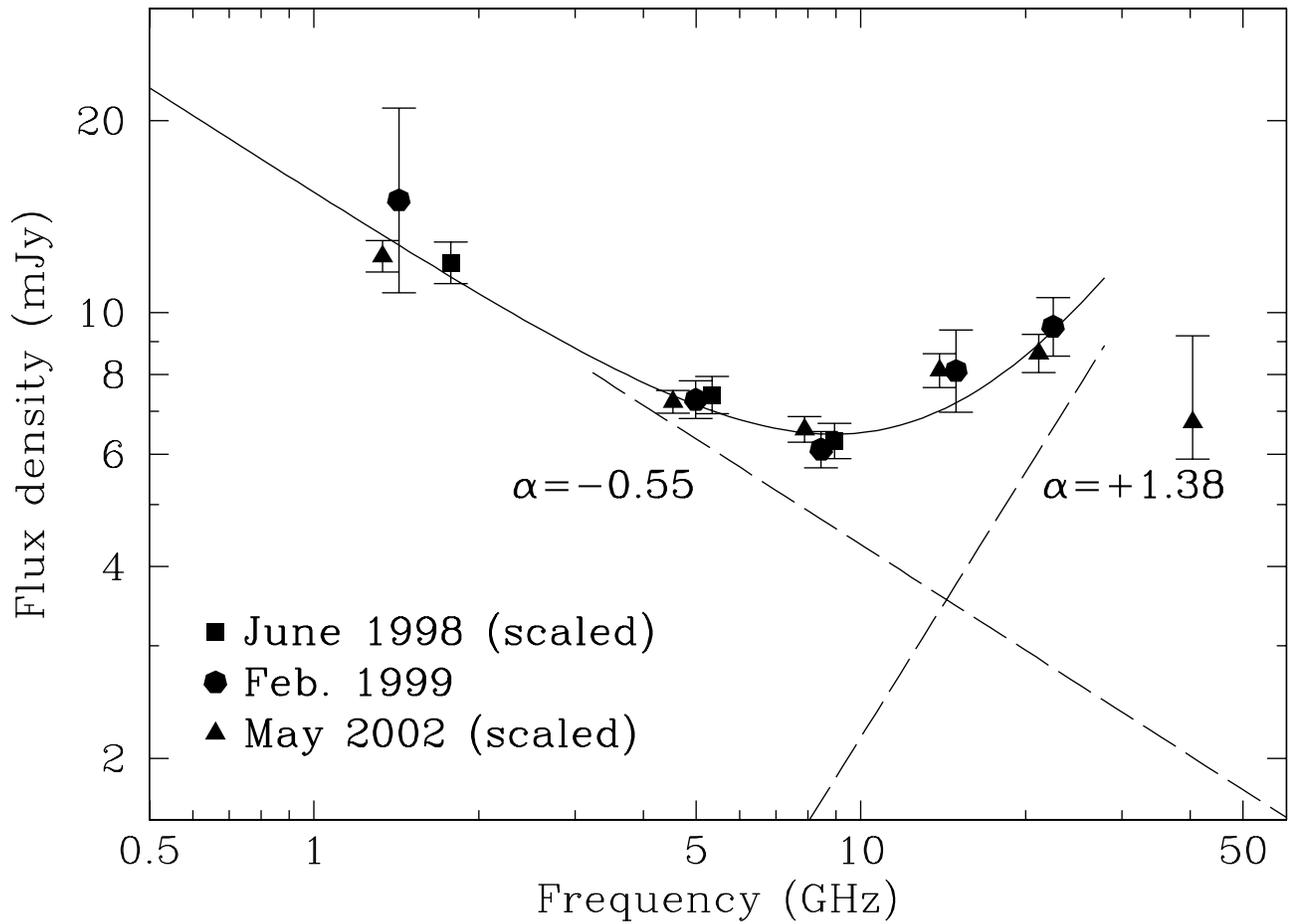}
\figcaption{The spectrum of SN~1986J as measured with the VLA\@.  Some
points have been shifted slightly in frequency for clarity.  The solid
line indicates a weighted, two-component fit to the spectrum up to
23~GHz and the flux density decay.  The measurements in 1998 and
2002 were scaled to epoch 1999 using the fit value of $\beta$ of
$-2.94$.  The dotted lines illustrate the two power-law components of
the fit spectrum, which have the indicated spectral indices,
$\alpha$. 
\label{spectrumf}}
\end{figure}

In Figure~\ref{spectrumf}, we plot our flux densities, all scaled with
the fit value of $\beta$ to the epoch of our last VLBI image.  We also
plot the two component fit to the time-averaged spectrum up to 23~GHz,
along with the normal and inverted components of the spectrum.  The
normal component has $S_{\rm 5\;GHz} = 6.3^{+0.6}_{-1.1}$~mJy and
$\alpha = -0.55^{+0.09}_{-0.16}$.  The spectral index of the normal
component is consistent within the uncertainties with the optically
thin spectral index of $\alpha = -0.67^{+0.08}_{-0.04}$ determined by
WPS90 for the period up to 1989. The inverted component has $S_{\rm
5\;GHz} = 0.8^{+1.0}_{-0.4}$~mJy and $\alpha = +1.4^{+0.6}_{-0.4}$.

The significant inversion in the spectrum visible above 10~GHz has not
been observed for any other supernova.  It is likely due to an element
of SN~1986J with an emission process very different from that of the
shell.  This inversion of the spectrum of
SN~1986J is seen on two different observing dates, using different
amplitude and phase calibrators.  Neither of the phase calibrators
shows any break in their spectrum (Fig.~\ref{vlacalf}).  We therefore
think that the inversion cannot be ascribed to any anomaly in the
calibration of the data.

Has the inversion of the spectrum been seen before in SN~1986J? 
An inversion of the spectrum was indeed seen earlier, in the fall of
1986, when the spectrum was unusually steep between 14.9 and 22.5~GHz,
with $\alpha = -1.5\pm 0.3$ (WPS90), but slightly inverted between 90
and 250~GHz \citep{TuffsCK1989}.
A subsequent measurement at 250~GHz made one year later by the same
authors, however, showed a much lower flux density and no inversion,
the spectrum between 5 and 250~GHz being consistent with a single
power-law with $\alpha \sim -0.63$ (WPS90).  WPS90 concluded that
either a transient component of SN~1986J was responsible for the
spectral inversion, or that the measurement by Tuffs et~al.\ (1989)
was in error.  We note, though, that for the second half of 1988,
WPS90 report a steep spectrum with $\alpha \simeq -0.67$ between 5 and
15~GHz, and a relatively flat spectrum, with $\alpha \sim -0.2 \pm
0.2$ between 15 and 23~GHz.  Such a flattening is compatible with the
presence of an inverted component of the spectrum.  Their data,
however, do not support an inverted component with as large a fraction
of the total flux density as we observed later.  Perhaps the inverted
component is highly variable.  If so, our finding that $\beta$ is
independent of frequency suggests that we fortuitously observed at a
time when the inverted component showed the same rate of decay as the
normal one.
Although our fit to the flux density decay and to the spectrum up to
23~GHz had a Chi-squared per degree of freedom of 0.5, indicating an
excellent fit, the spectral decomposition is likely not unique.  Our
measurement at 43~GHz suggests a high-frequency turnover in the
spectrum near $\sim 23$~GHz.
It is possible that the inverted component of the spectrum deviates
significantly from a power law at or even below 23~GHz.  At 23~GHz,
the likely minimum brightness temperature associated with the inverted
component of the spectrum is $3 \times 10^6$~K, derived by assuming a
size no larger than the size of the entire emission region in the
5~GHz VLBI image of 1999.1.

Can the inverted component of the spectrum be related to any
particular feature in our images of SN~1986J?  If the inverted
component of the spectrum is related to a compact feature in the
images, then the feature could be visible in our images.  Based on our
fit to the spectrum, we expect it to have $S_{\rm 5\;GHz} =
0.8^{+1.0}_{-0.4}$~mJy in 1999.1.  Since the rms of the background
brightness in our 1999.1 VLBI image (Fig.~\ref{hiresf}) is only
40~$\mu$\Jpb, such a feature should be visible.  We note, however,
that if the inverted component of the spectrum deviates from
a power law below 23~GHz, the value of $S_{\rm 5\;GHz}$ could be in
error.  The most obvious candidate for the feature in the images is
component C1, corresponding to the fit point source in our models,
which has a flux density of $\sim 1$~mJy.  Indeed, the compact
component, C1, is dominant in the VLBI image of 1999.1
(Fig.~\ref{hiresf}), but is not apparent in that of 1988.7.

\section{Astrometric Results\label{astroms}}

Phase referencing, which we did for our 1999.1 VLBI observations,
allows the determination of accurate relative positions. In principle,
the coordinates of the explosion center of a supernova can be
determined accurately with VLBI astrometry relative to a suitable,
stable, reference point nearby on the sky.  The core of a galaxy or a
quasar would be a suitable reference point: with present astrometric
techniques, these cores have no detectable proper motion
\citep[e.g.][]{Bartel+1986}, and can therefore be assumed for our
purposes to be stationary.  In particular, in the case of SN~1993J, we
used phase-referenced imaging with respect to the core of the nuclear
radio source of M81 to determine the explosion center of SN~1993J with
an accuracy of 45~\muas\ or 160~AU at the distance of the galaxy
\citep{BietenholzBR2001}. This measurement allowed us to relate the
expansion of the radio shell to the explosion center and to place
limits on any anisotropy in the expansion of the supernova with high
accuracy.

Our observations were made using the radio galaxy 3C~66A, which is only
42~arcmin away from SN~1986J, as a phase reference source. We display
an image of 3C~66A in Figure~\ref{3c66af}. The source is characterized
by a core component and a 20~mas long, one-sided jet to the
south-southwest. As the reference point for our astrometric
measurements, we used the position of the phase center of 3C~66A, which
is essentially that of the peak of the brightness distribution.  We
also use this position as the origin of the coordinate system of the
image in Figure~\ref{3c66af}.  The position of this point is the {\em
a priori}\/ position of 3C~66A assumed for the correlation of the data.
Its coordinates are R.A. = \Ra{2}{22}{39}{61148}, decl.\ =
\dec{+43}{02}{7}{7993} (J2000).  Relative to these coordinates, we
determined that the coordinates of the center of SN~1986J were R.A. =
\Ra{2}{22}{31}{3211} and decl.\ = \dec{+42}{19}{57}{2820} (J2000).

The accuracy of these relative coordinates is limited by errors in the
correlator model of the tropospheric and ionospheric delays, as well
as of UT1, polar motion, and antenna coordinates to roughly 100~\muas\
\citep[see][for an estimate of such error for the close pair of
quasars 3C~345 and NRAO~512]{Bartel+1986}.
Furthermore, the brightness peak is not necessarily located at the
core of the galaxy.  It may depend on possible structure changes near
the core \citep[see e.g.,][]{BietenholzBR2000, Bartel+1986}.  For
instance, the brightness distribution of the nuclear radio source of
M81 Bietenholz et~al.\ (2000) varies on short timescales, as does that
of many superluminal sources.  Also the brightness distribution of the
peak may be dependent on frequency and the size of the convolving
beam.  The brightness distribution of the central part of 3C~66A is
approximately Gaussian at our resolution and frequency, with a
major axis FWHM of $\sim 1$~mas.  We therefore assume a provisional
positional uncertainty of 1~mas in identifying the core or any related
other stationary point in 3C~66A\@.  Future observations, perhaps at
several frequencies, will likely allow a more accurate location of the
core.

These difficulties will have to be carefully considered in any future
astrometric observations of SN~1986J where 3C~66A is used as a
phase-reference source.  Despite the difficulties, such observations
have the potential to answer important questions concerning the
symmetry of the expansion, the effect of the CSM, and location of the
explosion center, and above all, the location and identification in
the images of the possible compact component responsible for the
spectral inversion.

\section{Size Determinations\label{sizes}}

\subsection{Size Measurements by Model-Fitting in the \uv~Plane\label{uvfits}}

In order to determine the expansion velocity and possible deceleration
of the radio source SN~1986J, we must estimate its size at each of our
epochs.  Given the complexity of the source and its changes with time,
such estimates are not as straightforward as in case of the symmetric
shell of SN~1993J.  We restrict ourselves here to average measures of
the shell radius.  We begin by discussing measures derived by fitting
geometrical models to the \uv\ (visibility) data.  Such model fits
have the advantage of being independent of the bias introduced by
convolving with a CLEAN beam, but they have the disadvantage of
being dependent on the model, which can at best only approximately
represent the complex structure of the supernova.

We fit the geometrical model by weighted
least-squares directly to the calibrated \uv~data.  The model we chose
consisted of two components.  The first component was the
two-dimensional projection of a three-dimensional, spherical,
optically thin shell of uniform volume emissivity.  Since the shell
was not sufficiently resolved to warrant also solving for its
thickness, we fix the ratio of the outer to inner angular radius at
$\thout / \thin = 1.25$, similar to that which provided the best fit
for the shell of SN~1993J \citep{Bartel+2002,
Bietenholz+2002}\footnote{For different adopted values of the ratio
$\thout / \thin$, the fit value of $\thout$ would change approximately
as the square root of $\thout / (1.25 \, \thin)$.  Provided, however,
that $\thout / \thin$ does not change during our observing interval,
the effect on $m$ is small, as is shown in \citet{Bartel+2002}.}.  The
second component was a point source, and was introduced to
parameterize the prominent hot-spot visible to the north-west in the
images (Figure~\ref{imagesf}; see also B91).  Both the position and
the flux density of the point source are free parameters.  From these
fits, we estimated \thout\ and give the results in Table~\ref{sizet}.

The statistical standard errors of \thout\ are generally $<20$~\muas\
or $<1$\%.  The true uncertainties of \thout, however, are almost
certainly dominated by systematic errors, in large part due to the
problem of fitting a simple geometric model to the complex and
evolving structure of SN~1986J\@.  Other systematic errors, e.g., those
due to the unknown shell thickness and to possible residual errors in
the amplitude and phase calibration also contribute to the overall
error budget \citep[see][for a longer discussion of the uncertainties
in a similar model-fitting process]{Bartel+2002}.  To estimate the
systematic errors we monitored the changes of the values of \thout\ at
various stages in the self-calibration process and for various
positions of nearby local minima in the $\chi^2$ space.  We obtained a
typical error of $\sim 70$~\muas. To estimate how much the values of
\thout\ depend on our choice of a suitable model for SN~1986J, we
modified our model by deleting the point source. We found that the
values of \thout\ changed by only $\sim 20$~\muas\ for the epoch
1988.7 and virtually not at all for epoch 1990.6. For epoch 1999.1 the
value changed by $\sim 160$~\muas. Such relatively large change,
coupled with a significant fit flux density of the point source, in
contrast to the results of the previous two epochs (see also
\S\ref{imagess}), indicates that the point source is an appropriate
addition to the shell model, at least for the 1999.1 epoch.  We varied
our model further and added a second point source for epoch 1988.7,
since a second strong southern compact component appears in the
high-resolution image of B91, and found that the value of \thout\
decreased by only $\sim 20$~\muas.  In conclusion we adopt a standard
error of 70~\muas\ for the values of \thout\ at each of our three
epochs.

\subsection{Size Measurements in the Image Plane \label{imexps}}

Since the images do not show detailed agreement with the simple
geometrical model, we also determined the size of the radio source
SN~1986J in the image plane and in a fashion not dependent upon the
assumption of a particular geometry.  Since our goal is to estimate
the expansion velocity and $m$, we need to use a consistent means of
determining the sizes at different epochs.  Measuring the size in the
image plane has the advantage of being able to more accurately account
for the complex structure of the shell, but the disadvantage that such
measurements are biased by the convolution with the CLEAN beam
\citep{Bartel+2002}.  In particular, a determination of the
deceleration parameter, $m$, is sensitive to the choice of the CLEAN
beam for each epoch.  For example, using a CLEAN beam whose size
increases linearly with time would bias the derived value of $m$
towards $m = 1$.  In general, any derived value of $m$ would be biased
toward the corresponding value $m_{\rm beam}$, where the size of the
CLEAN beam is $\propto t^{m_{\rm beam}}$.  To limit the bias, we chose
a dynamical CLEAN beam whose size evolves similarly to that of the
supernova. As mentioned already in \S\ref{imagess}, we used a CLEAN
beam whose FWHM evolves as $(t-t_0)^{0.75}$ in anticipation
of our later result for $m$ (see \S\ref{decels}) with $t_0 = 1982.7$
(WPS90).  We also used a CLEAN beam which evolved with $m = 0.70$ and
found no significant change in the results.

We derived two different kinds of radii from the images, the first
referred to the images' total flux density and the second to their
peak brightness.  The first kind of radius, \thfl, is the equivalent
radius of the area of the contour which contains 90\% of the total
flux density.  The second kind of radius, \thpk, is the equivalent
radius of the area of the 10\% contour of the images, that being the
lowest ``reliable'' contour in the 1999.1 image, which has the smallest
dynamic range.  We list the values for the two kinds of radii in
Table~\ref{sizet}.

Determining the exact uncertainties of \thfl\ and \thpk\ is not
straightforward, and as with the uncertainties of \thout, they are
likely dominated by systematic effects. However, they are probably
comparable to those of \thout. For instance, \thfl\ is only weakly
dependent on the size of the convolving beam: when we changed the size
of the beam by 10\% in each dimension, we found that \thfl\ changed
only by 3\%. We further investigated how phase self-calibration would
influence the measurement of \thfl\ for the case of the 1990.6 image:
we found that \thfl\ before and after phase self-calibration was
different by $<1$\%.  We thus think that the standard errors of \thfl\
and \thpk\ are indeed comparable to those of \thout.

\section{The Expansion Velocity, Explosion Date, and Deceleration \label{decels}}

We can compute the average expansion velocity directly from our
measurements of the average size.  We note that the interpretation of
the velocity so derived as a true expansion velocity rests on the
implicit assumption that the synchrotron emission observed at our
different epochs can be physically associated.  We first use the
radii, \thfl, which are the measure of the average outer radius of the
source least dependent on the source structure.  Taking a distance of
10~Mpc, we obtain an expansion velocity between 1990.5 and 1999.1 of
$6000 \pm 800$~\kms.

To determine the date of explosion, the possible deceleration, and the
velocity as a function of time, we again use the measurements of
\thfl.  In addition to the measurements for our three epochs, we used
earlier measurements of the size of SN~1986J\@.  Two such measurements
were reported by Bartel et~al.\ (1989).  They determined the
parameters of an elliptical Gaussian fit to VLBI data at 10.7~GHz in
1987.1 and at 5.0~GHz in 1987.4.  The geometric mean of the major and
minor axes FWHM values was $1.37 \pm 0.36$ and $1.26 \pm 0.24$~mas,
respectively.

In order to convert these measurements to values equivalent to \thfl\
we used the elliptical Gaussian fit to the data of epoch 1988.7 and
computed the geometric mean of the FWHM for its two axes. Assuming
that the ratio of \thfl\ and the geometric mean axis of the fit
Gaussian in 1987.1 and 1987.4 would be the same as it was in 1988.7,
we obtained two additional values that we used for the determination
of $t_0$ and $m$.  We list these values also in Table~\ref{sizet}.

\begin{figure}
\epsscale{1.0}
\plotone{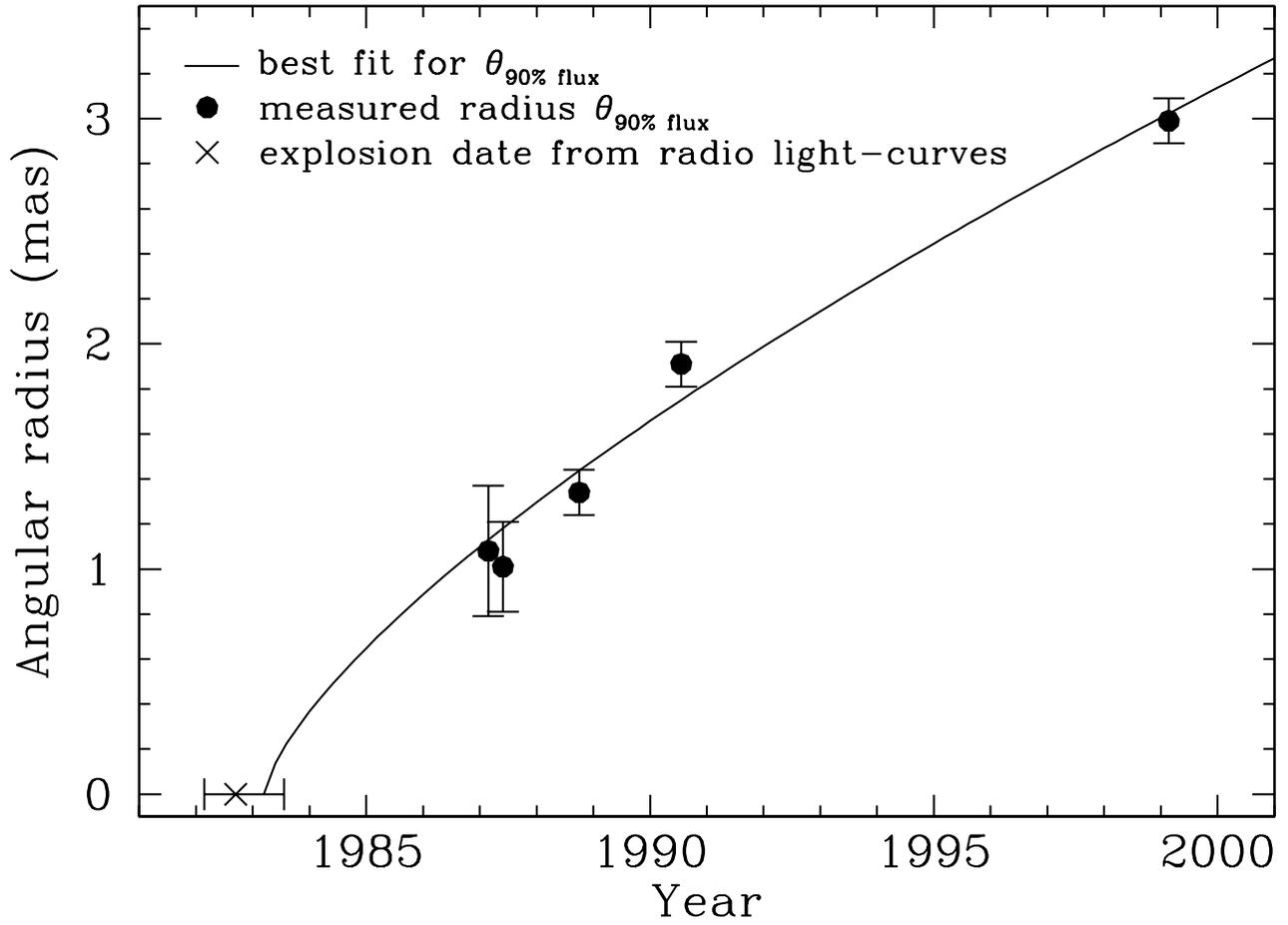} 
\figcaption{The expansion of SN~1986J as a function of time.  We plot
the angular radius, \thfl, against the year.  The line
indicates an expansion of the form $\theta \propto (t - t_0)^m$, with
$m = 0.71$ and $t_0 = 1983.2$, which are the values derived from our
best fit (see text \S\ref{decels}).
\label{expf}}
\end{figure}

We also added to our set of data points the estimate of the explosion
date, $t_0$, of $1982.7^{+0.8}_{-0.6}$ obtained from the radio light
curves (WPS90) and give it a weight comparable to those of our size
determinations.  We determine the values of $t_0$, $m$, and
$\theta_{\rm 1yr}$, the last being the angular radius of the supernova
after 1~year, by fitting the function of the form $\theta_{\rm
1yr}(t-t_0)^m$ to our values by weighted least squares.  We
plot the values of \thfl\ as well as the fit in Figure~\ref{expf}.
Since the fit function is non-linear, and the fit parameters may not
have Gaussian distributions, we derive the uncertainties from
Monte Carlo simulations, varying each value of \thfl\ and the estimate
of $t_0$ from the radio light curves according to their uncertainties,
and using 4000 random permutations.  This fit gives $t_0
=1983.2^{+1.4}_{-1.1}$, $m=0.71 \pm 0.11$, and $\theta_{\rm
1yr}=0.43^{+0.22}_{-0.13}$~mas.  These values correspond to an
expansion velocity at $t = 5.5$~yr (1988.7) of 8100~\kms\ and at $t =
15.9$~yr (1999.1) of 5900~\kms.

It is not clear how closely \thfl\ can be associated with the average
radius of the outer shock front.  If, for example, the source were
characterized by a pure shell morphology, then the radii, \thout, would
lead to a more direct estimate of the deceleration of the outer shock
front.  To explore our sensitivity to different definitions of
the outer radius, we also calculated the value of $m$ from our two
other kinds of radii, \thout\ and \thpk.

For the epochs 1987.1 and 1987.4, we estimated the values of \thout\
and \thpk\ in the same fashion as we estimated \thfl.  From the fit for
\thout, we obtained $m = 0.78 \pm 0.11$,
and for \thpk\ we obtained $m = 0.60 \pm 0.10$.  The explosion date
derived by WPS90 from the radio light curves is somewhat model
dependent.  If we omit their value from our fit to the values of
\thfl, we obtain $t_0 = 1986.2^{+0.7}_{-1.9}$, and $m =
0.45^{+0.15}_{-0.08}$.  Unfortunately, the VLBI size measurements
alone do not well constrain the fit, and the parameters $m$ and $t_0$
are highly correlated, so our individual uncertainties on each are
relatively large.  The fit value of $t_0$ corresponds to a time of
explosion somewhat, but not significantly, later than that of the
first radio detection in 1984.4 (Rupen et~al.\ 1987; van der Hulst
et~al.\ 1986).  Similarly fitting the values \thout\ and \thfl\ gives
$t_0=1984.7^{+1.9}_{-3.0}$, $m =0.62^{+0.24}_{-0.17}$ and $t_0 =
1986.1^{+1.2}_{-4.3}$, $m= 0.38^{+0.26}_{-0.10}$, respectively.

All our fits give explosion dates somewhat later than WPS90's value of
1982.7, suggesting that the true explosion date lies between that
value and the first radio detection in 1984.4.  All our values of $m$
indicate a moderate to strong deceleration.  The best average value
of $m$ is $0.71 \pm 0.11$, obtained from \thfl\ with the inclusion
of the estimate of $t_0$ from the light curves.

\section{The Evolution of the Magnetic Field\label{bfields}}

Using synchrotron radiation theory \citep{Pacholczyk1970}, we can
calculate the average magnetic field and infer its evolution.  We use
the measured radio light curve at 5.0~GHz, the spectral index of the
normal component of the spectrum, and a source radius, \rfl, which is
the linear radius corresponding to \thfl\ at distance $D = 10$~Mpc.
We also assume that the energy distribution of the ultra-relativistic
electrons in the interaction zone at any given time is a power law
with $n(E) \propto E^{-\gamma}$, where $E$ is the electron energy, and
$\gamma = 1-2\alpha$. It can then be shown \citep[e.g.][]{Bartel+2002}
that the magnetic field, $B$, in Gauss, is

$$B=\left(6 D^2 {\zeta_{B}\over{\zeta_{\rm rel} f r^3_{\rm 90\%\;flux}}}
    \right)^{2\over\gamma + 5} \left({{E_{\rm min}^{2-\gamma} S_{\nu}}
    \over {c_5 (\gamma -2)}}\right)^{2\over\gamma + 5}
    \left({\nu\over2c_1}\right)^{\gamma - 1\over \gamma+5},$$

\noindent with cgs units throughout.  Here $u_{\rm rel}=\zeta_{\rm rel} u_{\rm
e}$ and $u_{B}=\zeta_{B} u_{\rm e}$, where $u_{\rm e}$ is the internal
thermal particle energy density, $u_{B}$ the magnetic field energy
density with $u_{B} = {B^2\over8\pi}$, and $\zeta_{\rm rel}$,
$\zeta_{B}$ are taken to be constants.  Further, $f$, is the filling
factor which accounts for the hollow interior of the radio shell, any
irregularities in it, and for possible blocking of radiation from the
rear of the shell, $E_{\rm min}$ is the minimum energy of the
electrons, $c_1$ is a constant, and $c_5$ a function of $\gamma$
\citep[the last two given in][]{Pacholczyk1970}.

For a constant spectral index, $\alpha=-0.67$, ${\zeta_{B}\over
\zeta_{\rm rel}}=1$, $f = 0.4$, $E_{\rm min}$ equivalent to 0.511~MeV,
$\rfl = 2.0 \times 10^{17}$~cm for $t=5.5$ yr (1988.7), $\rfl \propto
t^m$, and $S_{\nu}\propto t^{\beta}$, we can compute $B(t)$. Taking
our value of $m$, which is relevant for the time interval 1983 to
1999, and taking $\beta=-2.22$, which was the average value between
1989 and 1999, and therefore best corresponds to our value of $m$, we
get $B(t) = 70 \; (t / 5.5 \, {\rm yr})^{-1.2}$~mG.

We observed no significant polarization of the radio emission, and
thus we cannot infer the geometry of the magnetic field from the
direction of the radio polarization.  However, the source size and the
above estimates of the magnetic field imply sufficient internal
Faraday rotation to completely depolarize the emission for any reasonable
density in the radio shell.

\section{Discussion \label{discuss}}

Three images of SN~1986J at consecutive epochs, 5.5, 7.4, and 15.9
years after the explosion have shown an expanding complex source that
strongly evolves with time.  SN~1986J is only the third supernova
after SN~1993J (see e.g.\ Bartel et al.\ 2000, Bietenholz et al.\
2001, for the latest sequence) and SN~1987A \citep{Gaensler+1997} for
which a sequence of images could be made\footnote{Sequences of images
were also made of the several-decades-old supernova remnant 43.31+592
and the possible supernova remnant 41.95+575 in the galaxy M82
\citep[][and references therein]{McDonald+2001}}.  The expansion is
decelerated.
Most interestingly, we found that the radio spectrum of SN~1986J is
inverted above 10~GHz from 1999 to 2002, a characteristic generally
absent in earlier radio spectra of SN~1986J, and not seen in any other
supernova. The rate of flux density decay has increased substantially
since 1989.  The evolution with time, the change in the spectrum, and
the change in the rate of flux density decay all clearly point to
non-self-similar evolution for this source In the remainder we will
discuss our findings in more detail.

\subsection{The Extended Structure and its Evolution with Time}

SN~1986J has a complex radio morphology that can be described as a
shell or composite, with hot spots modulating the ridge, and
protrusions piercing the rim, all evolving with time. These
characteristics contrast with those of the highly circular radio shell
of SN~1993J, but are similar to those of the compact source 41.95+575
in M82, which was believed to be a several-decades-old supernova
remnant with a distorted shell \citep{Bartel+1987, WilkinsondB1990}
although \citet{McDonald+2001} suggest a different origin.  The
deceleration is moderate to strong, with the exact value of the
deceleration parameter, $m$, depending on which definition of the
radius is taken to determine it.  The relatively large range from
$m=0.60$ to 0.78 obtained for different types of radii suggests that
even the evolution of the average radial profile of the supernova is
not self-similar, since if it were, all consistent measures of the
average radius should evolve with the same value of $m$.  Nonetheless,
all the values we obtained were consistent with our best-fit value of
$0.71 \pm 0.11$.

The radio light curve data suggest a strong change in the rate of flux
density decay between 1989 and 1999.  This change is likely to be
accompanied by changes in $m$, as is seen in SN~1993J
\citep{Bartel+2002}.  Our value of $m$ is the average value from the
explosion time till 1999.1.  It is possible, however, that $m$ was not
constant during this interval, since our measurements are also consistent
with an undecelerated expansion until $\sim 1992$, and a strong
deceleration after that.  Future size measurements are needed to
constrain the current deceleration and to determine whether it is
different from the average value.

Our best fit to the average deceleration, with $m = 0.71$, corresponds
to an expansion velocity in 1988.7 of $8100^{+1700}_{-830}$~\kms,
about half the velocity of the outermost protrusion given by
B91 after conversion to a distance of 10~Mpc.
In 1988.7 the protrusions extended to twice the radius of the radio
shell and had a mean velocity of 15,000~\kms.  Unfortunately, it is
not easy to track an individual protrusion between different epochs.
In 1988.7 the two most prominent protrusions were at position angles,
p.a., of $\sim -20\arcdeg$ and $\sim 110\arcdeg$. In 1990.6 the two
most prominent protrusions were oppositely directed, at p.a.\ $\sim
-20\arcdeg$ and $\sim 160\arcdeg$. However, in 1988.7, the south-east
protrusion is only sharply pointed at the 27.5\% contour. At lower
contours the protrusion is rather broad, extending to the p.a.\ $\sim
160\arcdeg$.  Perhaps it is the part of the protrusion near p.a.\ =
160\arcdeg\ that grows to the protrusion seen in 1990.6. In addition,
in 1990.6 a new protrusion appeared in the north-east. Perhaps the
evolution of protrusions was somewhat erratic.  In 1999.1 the
structure of the supernova is more circular and protrusions are less
prominent. They may have faded in emission or slowed. 

If the protrusions have faded more rapidly than the remainder of
the supernova, but still have the same relative extent that they did in
1989, then we may no longer be detecting the emission associated with
the outer shock front.  Consequently, the value of $m$ for the outer
shock would in fact be larger than our value of 0.71, which was
determined from average radii.

If the protrusions are slowing, for example because they are
encountering a denser CSM, then they must be more strongly decelerated
than the shell.  We can calculate their deceleration by fixing $t_0$
at 1983.2 and measuring the lengths of the protrusions from the origin
to the 90\% flux density contour in the images $a)$, $b)$, and $d)$ in
Figure~\ref{imagesf}, the last having the point source subtracted.  We
obtain a weighted mean $m$ for the protrusions of $\sim 0.6$. In this
case, the protrusions are currently being overtaken by the shock front.
Future measurements will show whether the deceleration of the shell
also increases to $\sim 0.6$.

What is the origin of the deviations from circular symmetry in
SN~1986J?  The radio emission is thought to be generated by the
interaction of the ejecta with the CSM which accelerates electrons to
relativistic energies and amplifies the magnetic field.  Anisotropies
and irregularities in the velocity pattern or density profile of the
ejecta, or in the density profile or magnetic field of the CSM could
lead to a non-circular shock front or to a modulation of the
brightness distribution along the ridge of the shell. In addition,
Rayleigh-Taylor (R-T) instabilities are expected to develop because of
the deceleration of the shock front, and also are also expected to
cause distortions of the radio structure.

It is not easy to distinguish between these different influences on
the radio structure of a supernova.  Highly asymmetric ejecta could be
caused by supersonic jets inducing the explosion of a supernova and
then propagating outward in both polar directions
\citep{Khokhlov+1999, PiranN1987}.  Recently \citet{McDonald+2001}
noted that the compact source in M82, 41.95+575, which previously was
similar to an elongated shell, now shows some characteristics of a
bipolar outflow, perhaps matching the jet scenario.  Other mechanisms
that could cause asymmetries in the structure of a supernova discussed
in the literature include the outbreak of plasma through weak points
of the shell \citep{Rees1987}, and giant loops of magnetic field lines
\citep{Shklovskii1981}.  In the case of SN~1993J, spectropolarimetry
gave evidence for an asymmetric density distribution in the ejecta
\citep{TrammellHW1993}.  Hydrodynamical studies, however, have
shown that the distortion of a supernova due to an axisymmetric
density distribution in the ejecta was much smaller than the
distortion for the complementary axisymmetric density distribution of
the CSM (Blondin et~al.\ 1996).

The complex structure could be largely due to influences of the
CSM\@. The model-fits to the radio light curves suggest a clumpy medium,
which could qualitatively explain some of the features of the radio
structure of SN~1986J\@.  \citet{BlondinLC1996} studied the interaction
of a supernova with an axisymmetric CSM density distribution having
density enhancement at the rotational equator of the progenitor, and
showed that such a distribution could form oppositely directed
protrusions with lengths two to four times the radius of the shell.
While we do see protrusions of this length, they do not seem to be
always oppositely directed.

The growth of R-T fingers may also influence the structure of
SN~1986J\@. Although the R-T fingers generally do not penetrate the
outer shock to form protrusions, \citet{JunJN1996} showed that the
interaction of vortices with a clumpy CSM allows some of the fingers
to grow beyond the outer shock.  Such fingers might then be disrupted
by the clumpy CSM and/or the conditions in the interaction region,
leading to their diminishing prominence.

\subsection{Synchrotron Self-Absorption at Early Times?}

It has been suggested that synchrotron self-absorption is the dominant
absorption mechanism early on in the evolution of a radio supernova
\citep[e.g.][]{Slysh1990}.  \citet{Chevalier1998} concluded on the
basis of then available data on SN~1986J that, while synchrotron
self-absorption was consistent with the low velocities determined from
the H$\alpha$ emission, it was inconsistent with the large sizes and
velocities determined through VLBI\@.  He determined that, for
synchrotron absorption to be significant, an outer radius of $< 3.2
\times 10^{16}$~cm was required at the peak of the 5-GHz light curve
in 1986.3. We confirm the VLBI sizes measured in 1988, and our
improved determination of the expansion velocity suggests a size in
1986.3 of 0.70~mas, corresponding to $1.1 \times 10^{17}$~cm, three
times larger than the maximum size for significant synchrotron
self-absorption.  The slow rise of the radio light curves is therefore
likely due to effects other than synchrotron self-absorption, such as
absorption within the emission region.  As WPS90 pointed out earlier,
such absorption would be consistent with filamentation of the radio
shell structure and at least qualitatively with the complexity of our
VLBI images.

\subsection{The Swept-Up Mass and its Implications}

The outer radius of SN~1986J in 1999.1 is 4.5\ex{17}~cm (calculated
from \thfl).  For a value $\dot M / w$ of $2.4 \times
10^{-4}$~\Msol~yr$^{-1}$ per 10~\kms\ inferred in part from the radio
light curves (WPS90) we calculate that the total swept-up mass in
1999.1 would be $\sim 3.4$~\Msol.

If we take the initial expansion velocity\footnote {For our best-fit
solution for $m$ and $t_0$, this is the velocity after three months,
and it is comparable to the early expansion velocity found for other
supernovae, e.g.\ SN~1993J \citep{Bartel+2000}.} to be $\sim
20,000$~\kms, then momentum conservation suggests that the mass of
the decelerated ejecta is $\sim 30$\% that of the material swept up by
1999.1, or $\sim 1$~\Msol.
We can then estimate the total kinetic energy of the interaction
region and the swept-up material to be $\sim 1.5$\ex{51}~ergs. The
total kinetic energy in SN~1986J consists of that of the swept-up
material, of the decelerated material, and of the still undecelerated
inner ejecta. The kinetic energy in the undecelerated ejecta is
probably $< 10^{51}$~ergs, giving a total kinetic energy in the range
of 1.5 to $\sim 2.5 \times 10^{51}$~ergs.  This is comparable to, or
possibly somewhat larger than, the total kinetic energy estimated for
a supernova explosion of a massive star, suggesting perhaps that the
average value $\dot M / w$ is lower than given above.  In fact,
assuming that $\dot M / w$ is constant in time is equivalent to
assuming $s=2$ (where $\rhoCSM \propto r^{-s}$). 

We can calculate $s$ within the context of the mini-shell model and
self-similar evolution.  In this model $s$ is related to the
observables $m$, $\beta$ and $\alpha$ by $\beta = -(3 - \alpha)(1 - s
m/2) + 6m - \alpha m$ (Chevalier 1982a, b; see also Fransson,
Lundqvist, \& Chevalier 1996).
Solving for $s$, we find $s = 2.6$.  We have shown that SN~1986J does
not evolve in a self-similar fashion, so the model will not apply in
detail.  It does, however, seem likely that the external density
profile is in fact steeper than $s = 2$, at least for $t > 5.5$~yr.  A
steeper profile is in agreement with the interpretation of recent
X-ray results \citep{Houck+1998}.  This would imply on average a lower
value of $\dot M /w$ than that determined early on by WPS90.  Perhaps
$\dot M / w$ was much smaller tens of thousands of years before the
explosion and increased to the high value given by WPS90 only shortly
before the star died.

\subsection{Interpretation of the Inversion in the Spectrum\label{taus}}

We found a clear inversion in the radio spectrum above 10~GHz
(Fig.~\ref{spectrumf}). The inverted part of the spectrum has a flux
density which rises with frequency at least up to $\sim 23$~GHz,
although there is likely a turnover or at least a flattening above
that frequency.

The inversion in the spectrum suggests two distinct populations of
relativistic electrons.  The first is that of shock-accelerated
electrons in the interaction region, which are responsible for the
normal, uninverted component of the spectrum.  The second is that
responsible for the inverted component of the spectrum, and whose
origin is unknown at this point.  Since the flux density above 10~GHz
is characterized by the same rate of flux density decay
as seen at lower frequencies, the mechanism responsible for the
inverted component of the spectrum is probably coupled to the
expansion of the supernova, which drives the time-dependence of the
normal component of the spectrum.  The inverted part of the spectrum
cannot then be due to a superposed HII region, which would not show any flux
density decay.  Accretion onto a compact object also often produces
flat or inverted spectra, albeit not usually at such high radio
luminosity for a stellar-mass object.  A compact object is expected
since the progenitor of SN~1986J was a massive star. Again, however,
it seems unlikely that accretion onto the compact object would show
the same time dependence as the emission from the expanding shell.

The inverted spectrum might be produced by either free-free or
synchrotron self-absorption.  Free-free absorption is characterized by
a spectral index, $\alpha = +2.5$ and synchrotron self-absorption by
$\alpha = +2.1$.  Both absorption mechanisms would show a steep rise
up to the frequency where the optical depth, $\tau_\nu$, is near
unity, and then a turnover, provided the unabsorbed emission is
characterized by $\alpha \lesssim +2$.  If the source evolves, then
$\tau_{\nu}$ and the spectrum are expected to change with time. In
particular, if the source expands, $\tau_{\nu}$ would be expected to
decrease, and the observed flux densities at frequencies where
$\tau_\nu > 0$ would be expected to increase relatively rapidly.  This
is in distinct contrast to our observation that the spectrum remained
unchanged within the errors between 1999 and 2002, and that the flux
density at all the observed frequencies decreased at the same rapid
rate.  This characteristic is puzzling and suggests that 1)
$\tau_{\nu}$ remains relatively constant or even increases with time,
and 2) the emission mechanism is coupled to the expansion of the
supernova.

What is the absorption process? If the inversion of the spectrum is
due to synchrotron self-absorption, we can compute the magnetic field,
$B$, of the self-absorbed source and compare it with the equipartition
field of the supernova.  In particular, for a source like C1 with a
FWHM $\lesssim 0.5$~mas, the flux density at 20~GHz of 5~mJy
implies a magnetic field $B\gtrsim 2 \times 10^5$~G\@.  Such a field is
seven orders of magnitude higher than the average equipartition field
and therefore unlikely to be realized in a hot spot of the expanding
shell. Given this characteristic coupled with points 1) and 2) above,
it is unlikely that the inversion of the spectrum is due to
synchrotron self-absorption.  Could the inversion then be due to
free-free absorption of the flat-spectrum emission of a pulsar nebula?

\subsection{A Pulsar Nebula?\label{pwns}}

Already in 1988, B91 noted that the complex morphology of SN~1986J was
not inconsistent with that of a shell with a weaker pulsar-wind nebula
(PWN) in its center.  We indicated in \S\ref{lightcurves} above how a
new, inverted component in the spectrum may be related to the compact
feature C1 in our latest image, both of which were detected in 1999
but not previously.

Could this feature be a PWN?  A PWN would only become visible once the
optical depth of the interior of the shell had decreased to $\lesssim
1$, either through expansion or through fragmentation of the ejecta
\citep{BandieraPS1983}.  Since the putative PWN would just now be
becoming visible, this suggests that the optical depth would just now have
decreased to $\sim 1$.  However, we argued in \S\ref{taus} above that
the optical depth is, if anything, increasing with time, rather then
decreasing as required by this interpretation.

If we nonetheless associate the image component C1 with the inverted
component of the spectrum and with a putative PWN, then the latter's
flux density at 5~GHz is $\sim 1$~mJy.  At 10~Mpc, this corresponds to
a spectral luminosity $100 \times$ that of the Crab Nebula.  In
agreement with the observed luminosity, \citet{BandieraPS1984}
calculated that a young PWN would have a luminosity 10 to 1000 times
that of the Crab Nebula.  They also calculated that the luminosity
would decay with time, again consistent with our observations,
although they do not find the decay to be as rapid as we have observed.

A PWN would be expected to be near the explosion center of SN~1986J,
since pulsar velocities, typically $< 500$~\kms, are much smaller than
our observed expansion velocity of $> 5000$~\kms.  The component, C1,
however, is located approximately half-way between the center and the
perimeter of the source.
It may be that the geometric center as determined either from
the shell fit or the center of, say, the contour containing 90\% of
the flux density, is not the center of the explosion.  The supernova's
structure is complex and the expansion may be anisotropic.  In
particular, the explosion center might be located close to C1.  Only
further phase-reference VLBI observations with accurate measurements
of the proper motions of SN~1986J could shed more light on this
possibility.

It may also be that a feature in the image closer to the geometric
center is a candidate for the PWN. We note that a south-western
extension of C1, labeled C1* on Figure~\ref{hiresf}, is much closer to
the geometric center, and has a brightness of $\sim 0.3$~mJy.
Again, further VLBI observations, preferably at high frequencies and
phase referenced to 3C~66A, are needed to test the nature of this
component.

\section{Conclusions\label{concs}}

\noindent Here we give a summary of our main conclusions.

\begin{trivlist}

\item{1.} The sequence of images of SN~1986J from 5.5 to 15.9 yr after
the explosion is only the second such sequence, after that of SN~1993J,
where the expansion and evolution is so clearly visible.

\item{2.} The sequence shows a complex source with a shell or composite
morphology and protrusions that distort the perimeter of the
supernova.

\item{3.} The structure of the supernova changes with time. The
protrusions diminish somewhat in prominence, the shell structure
becomes more visible, and a compact source with a flux density of
$\sim 1$~mJy emerges half way between the perimeter and geometric
center.

\item{4.}  The spectrum shows a significant inversion above 10~GHz,
which was not present before 1989.  If the spectrum is composed of two
power laws, the component with an inverted spectrum has a spectral
index of $\alpha = +1.4^{+0.6}_{-0.4}$ below 23~GHz.  The normal
component of the spectrum has $\alpha = -0.55^{+0.09}_{-0.16}$, which
is consistent with the value measured in 1989. Between 1998 to 2002,
the spectrum remains relatively unchanged.

\item{5.}  The inverted component of the spectrum or the image
component C1 may be related to a pulsar nebula, but the current
evidence is not conclusive.

\item{6.} The explosion date is estimated to be $1983.2 \pm 1.1$,
derived from both our expansion measurements and radio light curve
modeling of others.

\item{7.} The expansion is found to be moderately to strongly
decelerated, with the average outer radius being $\propto
t^{0.71\pm0.11}$ between $t = 0$ and 15.9~yr.  The expansion
velocity in 1999.1 was 6000~\kms, only about one third of the
extrapolated velocity after three months of $20,000$~\kms.  

\item{8.} With the assumptions given, the average magnetic field is
70~mG at $t=5.5$~yr and declines $\propto t^{-1.2}$ between $t = 5.5$
and 19.2~yr. 

\item{9.} The flux density decreases $\propto t^\beta$ with $\beta =
-2.94 \pm 0.24$ from $t = 15.9$ to 19.2~yr after the explosion.  The
rate of flux density decay increased substantially since $t = 5.5$~yr,
when $\beta$ was $-1.18^{+0.02}_{-0.04}$.

\item{10.} The changing structure, the emergence of an inverted part
of the spectrum, and the change in the rate of flux density decay all
clearly point out that the evolution of SN~1986J is not self-similar.
Its evolution cannot be adequately described by self-similar
solutions.

\end{trivlist}

\acknowledgements

ACKNOWLEDGMENTS. 

M. Craig helped with the reduction of the 1999 February VLBI data
during his term as summer student at York university in 2000.  We thank
Natural Resources Canada for helping with the observations at the
Algonquin Radio Observatory.  Research at York University was partly
supported by NSERC.  NRAO is operated under license by Associated
Universities, Inc., under cooperative agreement with NSF.  The
European VLBI Network is a joint facility of European and Chinese
radio astronomy institutes funded by their national research
councils. The NASA/JPL DSN is operated by JPL/Caltech, under contract
with NASA.  We have made use of NASA's Astrophysics Data System
Abstract Service.

\clearpage
\onecolumn

\clearpage

\begin{deluxetable}{l  c  l@{ }l@{ }l@{ }l@{ }l@{ }l@{ }l@{ }l@{ }l@{ }l@{ }l@{ 
}l@{ }l@{ }l@{ }l@{ }l@{ }l@{ }l@{ }l@{ }l@{\hspace{0.2in}} c c c c}
\tabletypesize{\scriptsize}
\rotate
\tablecaption{VLBI Observations of SN~1986J\label{obst}}
\tablewidth{570pt}
\tablehead{
& &\multicolumn{20}{c}{ANTENNA\tablenotemark{a}} & \colhead{Total} & 
\colhead{On-Source} &\colhead{Recording} &\colhead{Circular} \\
\colhead{Date}  & \colhead{Freq.}
 &\multicolumn{20}{c}{~}
 &\colhead{time\tablenotemark{b}} & \colhead{time\tablenotemark{c}} & 
\colhead{Mode\tablenotemark{d}} &\colhead{Polarization\tablenotemark{e}} \\
 & \colhead{(GHz)} & Eb & Mc & Nt & On & Go & Ro & Aq & Gb & Hs &  Y & Br & Fd & 
Hn & Kp & La & Mk & Nl & Ov & Pt & Sc
 &\colhead{(hr)}  &\colhead{(baseline-hr)} 
}
\startdata
 1988 Sep 29 & 8.4 & X &   &   & X & X & X &   & X & X & X &   &   &   &   &   &   &   & X*&  &    
 &   12.3  & \phn 71   & III-A & R \\
 1990 Jul 21 & 8.4 & X &   &   & X & X & X & X & X & X & X &   &   &   &   &   &   &   &   &  &    
 &   22.2  &  114   & III-A & R \\
 1999 Feb 22 & 5.0 &   & X & X & X &   &   &   &   &   & X & X & X & X & X & X & X & X & X & X & X & 
  11.8  & 266 &  256-8-2 & R+L \\
\enddata
\tablenotetext{a}{
  Ef= 100m, MPIfR, Effelsberg, Germany;\phn
  Mc=  32m, IdR-CNR, Medicina, Italy;\phn
  Nt=  32m, IdR-CNR, Noto, Italy;\phn
  On=  20m, Onsala Space Observatory, Sweden;\phn 
  Go=  70m, NASA-JPL, Goldstone, CA, USA;\phn
  Ro=  70m, NASA-JPL, Robledo, Spain;\phn
  Aq=  46m, ISTS (now CRESTech/York Univ.), Algonquin Park, Ontario, Canada;\phn
  Gb=  43m, NRAO, Green Bank, WV, USA;\phn
  Hs=  36m, Haystack Observatory, Westford, Ma, USA;\phn
  Y = equivalent diameter 130m, NRAO, near Socorro, NM, USA;\phn
  Br=  25m, NRAO, Brewster, WA, USA;
  Fd=  25m, NRAO, Fort Davis, TX, USA;
  Hn=  25m, NRAO, Hancock, NH, USA;
  Kp=  25m, NRAO, Kitt Peak, AZ, USA;
  La=  25m, NRAO, Los Alamos, NM, USA;
  Mk=  25m, NRAO, Mauna Kea, HI, USA;
  Nl=  25m, NRAO, North Liberty, IA, USA;
  Ov=  25m, NRAO, Owens Valley, CA, USA (asterisk denotes use of the 40m antenna 
at this
location);
  Pt=  25m, NRAO, Pie Town, NM, USA;
  Sc=  25m, NRAO, St. Croix, Virgin Islands, USA.
  }
\tablenotetext{b}{Maximum span in hour angle at any one antenna.}
\tablenotetext{c}{Number of baseline-hours spent on SN~1986J, after data 
calibration and editing.}
\tablenotetext{d}{Recording mode: III-A = Mk III mode A, 56~MHz
recorded (effectively only 48~MHz at the VLA); \\
256-8-2 = VLBA format, 256 Mbps recorded in 8 baseband channels with 2-bit 
sampling.}
\tablenotetext{e}{The sense of circular polarization recorded: R = right
and L = left circular polarization (IEEE convention).}

\end{deluxetable}

\begin{deluxetable}{l@{\extracolsep{45pt}} l l}
\newcommand{\spquote}{\multicolumn{1}{c}{\"}}
\tablecaption{Total Flux Densities Measured at the VLA\label{vlafluxt}}
\tablewidth{0pt}
\tablehead{ 
\colhead{Date} & 
\colhead{Freq.}  
& \colhead{Flux Density\tablenotemark{a}} \\
& \colhead{(GHz)} & \colhead{(mJy)}
}
\startdata
1998 June 5    & \phn 1.67 &     $13.7 \pm 0.9$ \\
 \spquote      & \phn 4.99 & \phn $8.5 \pm 0.5$ \\
 \spquote      & \phn 8.41 & \phn $7.2 \pm 0.4$ \\ 
1999 Feb.\ 22  & \phn 1.51 &     $15.0 \pm 5.0$ \\
 \spquote      & \phn 4.99 & \phn $7.3 \pm 0.5$ \\
 \spquote      & \phn 8.49 & \phn $6.1 \pm 0.4$ \\
 \spquote      &     15.01 & \phn $8.1 \pm 1.2$ \\
 \spquote      &     22.54 & \phn $9.5 \pm 1.0$ \\
2002 May 25    & \phn 1.51 & \phn $7.1 \pm 0.7$ \\
 \spquote      & \phn 4.94 & \phn $4.2 \pm 0.3$ \\
 \spquote      & \phn 8.49 & \phn $3.8 \pm 0.3$ \\
 \spquote      &     15.01 & \phn $7.1 \pm 0.7$ \\
 \spquote      &     22.54 & \phn $5.0 \pm 0.6$ \\
 \spquote      &     43.36 & \phn $3.9\;^{+\;2.1}_{-\;0.9} $ \\
\enddata
\tablenotetext{a}{The VLA flux densities are derived using data
self-calibrated in phase, from both image-plane measurements and
model-fits to the visibility data, accounting for the presence of
disk-emission from the galaxy at the lower frequencies.  The
standard errors include statistical and systematic contributions.}
\end{deluxetable}

\begin{deluxetable}{l r r c c c}
\tablecaption{Angular Radii for SN~1986J \label{sizet}}
\tablewidth{0pt}
\tablehead{ 
\colhead{Date} & \colhead{Age\tablenotemark{a}} & \colhead{Freq.} & 
                 \colhead{Model-fit radius} &
                 \multicolumn{2}{c}{Image-plane radii} \\
 & & & \colhead{\thout\tablenotemark{b}}
     & \colhead{\thfl\tablenotemark{c}}
     & \colhead{\thpk\tablenotemark{d}} \\
     & (yr) & (GHz) & (mas) & (mas) & (mas)
}
\startdata
1988 Sep. 29 &  5.55 &  8.4 &$1.05\pm0.07$ &$1.34\pm0.10$& $1.66\pm0.12$ \\
1990 July 21 &  7.35 &  8.4 &$1.34\pm0.07$ &$1.91\pm0.10$& $2.13\pm0.12$ \\
1999 Feb. 22 & 15.94 &  5.0 &$2.33\pm0.07$ &$2.99\pm0.10$& $3.14\pm0.12$ \\[10pt]
1987 Feb. 23\tablenotemark{e} 
             &  3.95 & 10.7 &$0.86\pm0.23$ &$1.08\pm0.29$& $1.33\pm0.36$ \\
1987 May  30\tablenotemark{e} 
             &  4.21 &  5.0 &$0.80\pm0.15$ &$1.01\pm0.20$& $1.27\pm0.24$ \\
\enddata 
\tablenotetext{~}{The listed uncertainties are approximately standard
errors (see text).}
\tablenotetext{a}{Calculated by assuming our best-fit
explosion date of 1983.2.}
\tablenotetext{b}{The outer angular radius, \thout, of the
shell model fit to the \uv~data.  The shell model consisted of the
projection of a spherical shell of uniform volume emissivity and
an additional point source to parameterize the most prominent
hot-spot.}
\tablenotetext{c}{The equivalent angular radius, \thfl\ ($= \sqrt{{\rm
area}/\pi}$) of the contour which contains 90\% of the total flux
density in the image.  All three images were convolved with a beam
whose FWHM evolves as $t^{0.75}$.}
\tablenotetext{d}{The equivalent angular radius \thpk\ ($= \sqrt{{\rm
area}/\pi}$) of the contour at 10\% of the peak brightness. All three
images were convolved with a beam whose FWHM evolves as
$t^{0.75}$.}
\tablenotetext{e}{These values were calculated from FWHM radii of the fit
elliptical Gaussian taken from Bartel et~al.\ (1989); see text
\S\ref{imagess}.}

\end{deluxetable}

\clearpage

\epsscale{0.41}
\begin{figure}
\plotone{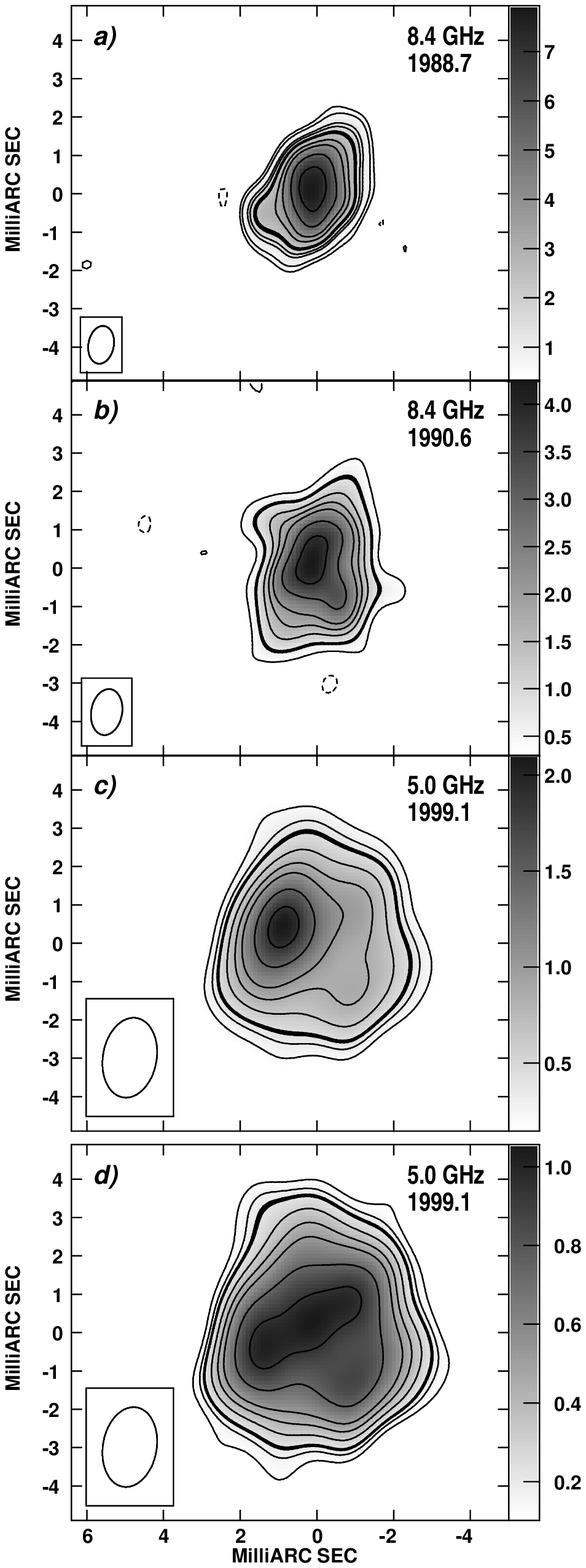}
\end{figure}
\begin{figure}
\figcaption{Four images of SN~1986J\@.  The epoch and frequency are
given in the upper right corner of each panel.  The first three are
the images of SN~1986J at our three VLBI epochs.  The fourth, shown
for comparison, is the image in 1999.1 with the fit point source
subtracted (see text, \S\ref{uvfits}).
The FWHM of the convolving beam, indicated at lower left in each
panel, was chosen to evolve as $(t - 1982.7)^{0.75}$. In all four
panels, the lowest contour is drawn at $3\times$ the rms of the
background brightness, and the heavier contour is the one which
contains 90\% of the total flux density in the image (see text,
\S\ref{imexps}). For all images, north is up, and east is to the left.
\newline $a$) 1988.7: the FWHM of the CLEAN beam was 1.0~mas
$\times$ 0.66~mas at $-10$\arcdeg; the contours are drawn at $-5$, 5,
8, 16, {\bf 27.5}, 40, 50, 70, and 90\% of the peak brightness of
7.9~m\Jpb.
\newline $b$) 1990.5: the FWHM of the CLEAN beam was 1.2~mas
$\times$ 0.81~mas at $-10$\arcdeg; the contours are drawn at $-8$, 8,
{\bf 17.6}, 30, 40, 50, 70, and 90\% of the peak brightness of
4.2~m\Jpb.
\newline $c$) 1999.1: the FWHM of the CLEAN beam was 2.1~mas
$\times$ 1.4~mas at $-10$\arcdeg; the contours are drawn at 5, 8,
{\bf 12.1}, 30, \dots\ 90\% of the peak brightness of 2.1~m\Jpb.
\newline $d$) 1999.1: The CLEAN beam was as in $c)$.  The contours are
drawn at 11, 16, {\bf 20.7}, 30, \dots\ 90\% of the peak brightness of
1.0~m\Jpb.  Note that the heavier contour is that which contains 90\%
of the flux density remaining {\em after}\/ the point source has been
subtracted.
\label{imagesf}}
\end{figure}

\begin{figure}
\epsscale{0.50}
\plotone{3c66-big.eps}  
\figcaption{A naturally weighted image of the source 3C~66A, which was
used as a phase-reference source.  The origin is at the phase center
of 3C~66A: R.A. = \Ra{2}{22}{39}{61148}, decl.\ =
\dec{+43}{02}{7}{7993} (J2000).  Contours are drawn at 0.5, 1, 2, 4,
8, 15, 30, {\bf 50}, 70, and 90\% of the peak brightness of 0.69~\Jpb,
with the 50\% contour being emphasized.  The FWHM of the CLEAN beam,
indicated at lower left, was 3.8~mas $\times$ 3.5~mas at p.a.\
$= -11$\arcdeg. North is up, and east is to the left.
\label{3c66af}}
\end{figure}

\end{document}